\documentclass[manuscript, screen, nonacm]{acmart} 

\usepackage{booktabs}
\usepackage{tabularx}
\usepackage{longtable}
\usepackage[dvipsnames]{xcolor}

\graphicspath{{figures/}}

\newcommand{\HH}[1]{\textbf{H#1}}

\newcommand{\spara}[1]{\smallskip\noindent\textbf{#1}}
\newcommand{\subspara}[1]{\smallskip\noindent\emph{#1}}

\usepackage{ragged2e}
\usepackage{pifont}
\newcommand{\realdot}{\raisebox{0.15ex}{\setlength{\unitlength}{0.8pt}\begin{picture}(6,6)\put(3,3){\circle*{6}}\end{picture}}}
\newcommand{\invdot}{\raisebox{0.15ex}{\setlength{\unitlength}{0.8pt}\begin{picture}(6,6)\thicklines\put(3,3){\circle{6}}\end{picture}}}
\newcommand{\cmark}{\ding{51}}
\newcommand{\xmark}{\ding{55}}

\begin{document}

\title{The Popular Get More Popular: Auditing Brand Bias in Generative Search}

\author{Giorgio Monaco}
\email{giorgio.monaco@mail.polimi.it}
\affiliation{\institution{Politecnico di Milano}\city{Milan}\country{Italy}}
\author{Jacopo Lenti}
\email{jacopo.lenti@polimi.com}
\affiliation{\institution{Politecnico di Milano}\city{Milan}\country{Italy}}
\author{Francesco Pierri}
\email{francesco.pierri@polimi.it}
\affiliation{\institution{Politecnico di Milano}\city{Milan}\country{Italy}}

\raggedbottom

\begin{abstract}
The shift from traditional to generative search has fundamentally transformed online product discovery. 
Because conversational recommendations synthesize choices into a few direct answers, subtle systemic preferences toward specific providers can exert profound commercial impacts. 
In this work, we audit this phenomenon through two distinct dimensions: (i) \textit{brand popularity bias}, the systematic preference for market-dominant entities over long-tail alternatives, and (ii) \textit{brand familiarity bias}, the ranking advantage an entity encoded in an LLM's parametric memory holds over an unfamiliar competitor. 
First, evaluating commercial queries across major engines, we observe persistent brand popularity bias: dominant brands in the top quintile are cited $1.4\times$ to $2.8\times$ more frequently than the average brand within a sample of 1,176 top market players, already present in parametric memory that external web retrieval fails to mitigate. 
Second, through controlled in-context experiments holding product attributes constant, we find that substituting authentic brand signatures with unfamiliar ones significantly degrades recommendation rank. 
Conversely, fact-preserving content rewrites yield negligible ranking gains. 
Taken together, our findings demonstrate that generative search engines reinforce a ``rich-get-richer'' dynamic by favoring established market leaders and imposing a parametric barrier to entry that standard content optimization cannot overcome.

\end{abstract}



\maketitle

\section{Introduction}
\label{sec:intro}
Chatbots and search engines powered by Large Language Models (LLMs) have shifted consumer product discovery from static URL ranking to conversational, fine-grained recommendations with synthesized justifications. 
Because users ascribe high credibility to these outputs and increasingly rely on them for everyday decisions~\citep{spatharioti2025effects}, generative search exerts substantial influence over consumer choice and market exposure. 
Auditing the mechanisms governing LLM-driven product recommendations is therefore of paramount importance.

Because these instruments operate at a global scale, minor systematic biases in their generation logic can produce profound market effects, as brands prioritized by generative search engines might gain distinct commercial advantages. 
While model preferences may sometimes reflect genuine product superiority, it is critical to disentangle meritocratic selection from systemic model bias. 
Trained on massive Web corpora, LLMs inevitably inherit and amplify pre-existing statistical biases \citep{gallegos2024bias}. 
While recent literature has audited LLMs for sociodemographic biases---including gender \citep{armstrong2024silicon}, ethnicity \citep{salinas2024name}, and social group \citep{kondapally2026sidebyside}---as well as prompting biases like persona \citep{jack2026persona} and cognitive biases \citep{filandrianos2025bias}, brand-level economic biases remain underexplored.

In this study, we formalize and audit two distinct dimensions of brand bias:
\begin{enumerate}
    \item \textbf{Brand Popularity Bias:} The systematic preference of LLMs toward highly popular, high-market-share brands over long-tail or niche alternatives. 
    Overrepresented in pre-training corpora, popular brands might enjoy a structural head start in user recommendations, which may create a ``rich-get-richer'' (or Matthew) effect~\citep{merton1968matthew}. This could reduce the visibility of local or specialized competitors~\citep{klimashevskaia2024survey,mansoury2020feedback}.
    \item \textbf{Brand Familiarity Bias:} The advantage a brand recognized within an LLM's parametric memory holds, at equal content, over an unfamiliar or novel entrant. 
    This dimension evaluates the algorithmic barrier to entry for newly emerging brands that lack pre-existing weight representations in frozen models.
\end{enumerate}

Concurrently, the shift toward generative search has spurred novel visibility-seeking strategies known as Generative Engine Optimization (GEO) \citep{aggarwal2024geo}.
Analogous to traditional marketing, GEO aims to alter page representations---adjusting style, tone, and keyword alignment---to maximize the probability of selection by LLM-based chatbots and search engines \citep{aggarwal2024geo,puerto2025cseo}.
However, the relative weight that LLMs assign to surface-level text optimization versus deeply memorized brand entities remains unclear. 
In this work, we contrast the impact of changing brand identity against rewriting content via GEO techniques to test whether LLMs decouple brand name authority from underlying product descriptions.

Notably, the normative relationship between brand bias and recommender fairness is nuanced \citep{burke2017multisided,wang2023fairness}. 
On one hand, fair recommendation engines should evaluate product merit independently of brand prominence, benefiting both niche vendors through equal exposure and consumers through increased diversity. 
On the other hand, established brands often correlate with higher consumer trust, quality control, and widespread user satisfaction. Prioritizing well-known brands may therefore align with average consumer expectations. 
Nevertheless, LLMs are known to suffer from mode collapse \citep{zhang2025verbalized}; in commercial recommendations, this behavior risks creating digital market oligopolies, suppressing fair competition, and locking users into homogeneous recommendations.
To understand these dynamics, we address two research questions:
\begin{itemize}
    \item \textbf{RQ1:} How do deployed generative search engines distribute visibility among brands across varying levels of popularity?
    \item \textbf{RQ2:} How do manipulations of a product page, such as swapping brands/products and using GEO rewriting strategies, affect its recommendation rank?
\end{itemize}

We evaluate these questions across four distinct commercial sectors in the Italian market: \texttt{Sport}, \texttt{Fashion\&Beauty}, \texttt{Insurance}, and \texttt{Travel}. 
In \textbf{Study 1}, we quantify brand popularity bias by evaluating brand text mentions and web source citations under both \texttt{search-on} (RAG-enabled) and \texttt{search-off} (parametric memory only) conditions. 
In \textbf{Study 2}, we evaluate brand familiarity bias and GEO influence by manipulating product descriptions---swapping real and invented brand names and applying GEO stylistic rewrites to test ranking stability.

Our contributions are the following:
\begin{itemize}
    \item We propose an experimental methodology to isolate and evaluate brand popularity bias and brand familiarity bias in generative search, accounting for web grounding, buyer personas, brand names, and GEO rewrites. The framework rests on a simple fairness criterion, equal rank at equal content, and in principle applies to any domain and model.
    \item We demonstrate that commercial generative search engines exhibit significant favoritism toward popular brands across both parametric generation and web-retrieved citations.
    \item We show that the same page ranks lower under an invented brand than under its real brand, at equal content (brand familiarity bias), and that GEO rewrites do not close this gap.
\end{itemize}

\section{Related work}
\label{sec:related}

Auditing algorithmic systems is a cornerstone of socio-technical inquiry, tracing back to foundational frameworks in traditional search engines~\citep{introna2000shaping, sandvig2014auditing, metaxa2021auditing}. 
However, the paradigm shift from static ranked lists of URLs to synthesized natural language answers has introduced novel systemic risks, prompting urgent calls for updated auditing paradigms suited for generative search~\citep{kirsten2026characterizing}.

\spara{Auditing Generative Search Engines.}
Generative search engines reconfigure the information ecosystem, impacting content creators, end users, model providers, and digital markets simultaneously~\citep{sadeddine2025large}. 
While generative search engines draw on a wider range of sources compared to traditional web search~\citep{kirsten2026characterizing}, their reliability remains deeply problematic. 
Grounding evaluations reveal that many generative claims are unsupported by or misattributed to their cited sources~\citep{liu2023verifiability}. 
Moreover, generative search exhibits persistent socio-technical vulnerabilities, including overconfidence, unannounced source selection heuristics, and high response stochasticity~\citep{venkit2025search, chen2026koudas, liu2023verifiability, schulte2026measure}. 
These dynamics obscure accountability, making it difficult for users to evaluate the factual integrity or commercial neutrality of recommended content.

\spara{Economic Fairness and Popularity Bias in Recommender Systems.}
Recommender system fairness was a critical area of study well before the advent of LLMs~\citep{biega2018equity}. 
Fairness in recommendation is inherently multi-sided, encompassing both \textit{consumer fairness} (ensuring users receive honest, diverse, and unbiased recommendations) and \textit{producer fairness} (ensuring items or vendors are not penalized for features unrelated to product merit)~\citep{burke2017multisided, singh2018exposure}. 
%
Focusing on generative models, LLMs inherit statistical distribution from their massive web pre-training corpora~\citep{yu2023large}, thus amplifying their bias to end users via authoritative and credible responses in natural language~\citep{sadeddine2025large}, thus mining model fairness and accountability.
Algorithmic biases in LLMs span social and attribute dimensions (favoring protected demographic groups), linguistic and knowledge dimensions (disadvantaging specific languages), system and optimization dimensions (driven by interface design), and data or propensity dimensions~\citep{ma2026fairness}.
Among data-driven biases, \textit{popularity bias} is particularly prevalent in recommender systems~\citep{klimashevskaia2024survey}. 
LLMs demonstrate a strong structural propensity to recommend head-items overrepresented in their training data~\citep{dong2026knowledge, chu2026incumbent}. 
Mitigating popularity bias directly enhances producer fairness toward long-tail and underrepresented entities~\citep{deldjoo2024fairness, dinnissen2023amplifying}. 
Under the broader umbrella of \textit{brand bias}, LLMs systematically prioritize global brands over regional or local alternatives~\citep{kamruzzaman2024brand}. 

\spara{Entity-Level, Name, and Geographic Preference Biases.}
LLM recommendation outcomes are also heavily modulated by prompt-level conditions and textual entity names. 
For instance, conditioning LLMs with distinct buyer personas significantly alters the commercial brands recommended in generated outputs~\citep{jack2026persona}.
Furthermore, changing only an entity's name reveals deep-seated model preferences. Audit literature in human audit studies established that audit testing via resume name manipulation reveals systemic racial and ethnic disparities, favoring names associated with White individuals~\citep{bertrand2004emily, gaddis2017names}. 
Parallel name-based biases persist in LLMs, where decision-support models penalize names tied to women and racial minorities~\citep{salinas2024name, armstrong2024silicon}. 
In product recommendation contexts, an item's product name and document ordering substantially distort its generated ranking~\citep{pfrommer2024ranking}.
Similarly, cross-cultural audits indicate that open-weight models originating from the US, Europe, and China exhibit geographic skews toward domestic brand origins when ranking shopping items, though explicitly stating the user's location can recalibrate these rankings~\citep{betello2026passport}.

\spara{Generative Engine Optimization (GEO) and Strategic Manipulation.}
The susceptibility of LLMs to textual manipulation introduces new surface vectors for commercial exploitation. 
Third parties can intentionally alter product page representations to inflate recommendation probabilities~\citep{filandrianos2025bias}. 
Mirroring human purchasing heuristics, LLMs exhibit vulnerability to embedded cognitive bias triggers—such as social proof, exclusivity, scarcity, discount framing, and authority claims—which significantly increase an item's generated visibility.
This vulnerability has driven the rise of \textit{Generative Engine Optimization (GEO)}, wherein web pages are strategically rewritten using statistics, quotes, and authoritative formatting to maximize AI retriever selection~\citep{aggarwal2024geo}. 
While some studies confirm the efficacy of GEO techniques, others observe mixed or neutral impacts~\citep{kumar2024manipulating, puerto2025cseo}. 
Crucially, the proliferation of strategic GEO manipulation creates a classic \textit{Prisoner's Dilemma} for content providers and vendors: individual entities have a strong competitive incentive to adversarially optimize their content for LLMs, regardless of competitor actions, ultimately degrading the overall utility and trustworthiness of the search ecosystem~\citep{nestaas2025adversarial}.
Because post-hoc data filtering is inherently complex, current mitigation strategies primarily target output generation via fairness-guided prompting~\citep{ma2026fairness}, logit re-confidencing~\citep{chen2024reconfidencing}, or in-context learning constraints~\citep{schubert2024incontext}.

\spara{Position of Our Contribution.}
Building upon the established literature in recommender fairness and generative search audits, this work disentangles brand bias along two distinct socio-technical axes: \textit{brand popularity bias} and \textit{brand familiarity bias}. 
While prior work has recognized general head-item dominance~\citep{chu2026incumbent, klimashevskaia2024survey}, we systematically measure how popularity bias restricts visibility for long-tail vendors across both inline textual mentions and web source citations. 
The same approach could be applied to other entities whose name a model may know, such as news outlets or other sources~\citep{dai2025media, khan2026agents}.
By explicitly evaluating models under both Retrieval-Augmented Generation (\texttt{search-on}) and non-retrieval (\texttt{search-off}) settings, we isolate the inherent bias of parametric model memory from grounding retrieval mechanisms. 

Furthermore, our study extends current auditing paradigms for RAG systems~\citep{liu2023verifiability, kirsten2026characterizing} by examining the initial search queries generated by LLMs.
\citet{chen2026koudas} observe that brands the model already knows are written into the search queries; we measure how this changes with brand popularity.
In some contexts, we reveal that popular brands secure a structural advantage early in the retrieval pipeline, shaping the initial web queries before final synthesis occurs. 
Finally, whereas existing entity-manipulation studies focus primarily on isolated name swaps or stylistic rewrites~\citep{aggarwal2024geo, pfrommer2024ranking}, our findings suggest that LLMs may evaluate brand identity in tandem with product-description coherence. 
We show that unfamiliar brands face a notable entry barrier even when paired with identical product specifications, while mismatched brand-product associations appear to incur an additional coherence penalty.


\section{Methods}
\label{sec:methods}

\begin{figure}[t]
  \centering
  \includegraphics[width=0.95\linewidth]{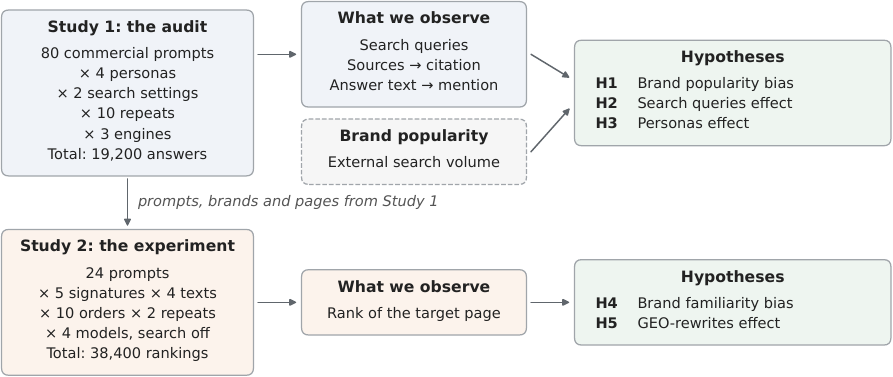}
  \Description{Flow diagram of the two studies. Study 1 sends commercial prompts with and without personas and search to three engines and keeps search queries, sources and answer text. Study 2 has four models rank five product pages while one page changes its signature or its text.}
  \caption{Study~1 is an audit of three deployed generative search engines on commercial prompts, with \texttt{search-on} and \texttt{search-off}, and with and without a persona context. Study~2 takes 24 of these prompts and manipulates two features of one product page, the brand that signs it and its text (GEO rewrites), while four models rank it among five pages with \texttt{search-off}.}
  \label{fig:pipeline}
\end{figure}

Figure~\ref{fig:pipeline} shows the pipeline of the two studies. 
Study~1 is an observational, data-driven audit of three deployed engines. We collect their answers to a set of real commercial prompts and measure how the popularity of a brand relates to its mention share and citation share (\textit{brand popularity bias}).
Study~2 is instead a controlled experiment, where we isolate the effect of brand name swap (\textit{brand familiarity bias}) and strategic rewrite (\textit{GEO effect}) on the rank of a page.
Table~\ref{tab:engines} lists the engines of both studies.

\subsection{Study 1: the audit}
\label{sec:m_s1}
We study brand popularity bias, by interrogating the LLMs under different settings, investigating the mention and citation shares of brands, paired with brand popularity.

\spara{Data collection.} 
The prompts are recommendation questions derived from what real users search for. 

We collected keyword data from DataForSEO for four sectors, \texttt{Insurance}, \texttt{Travel}, \texttt{Fashion\&Beauty}, and \texttt{Sport}, keeping commercial keywords that name no brand. 
In each sector, we drew keywords at random across levels of search volume, with at most two per sub-topic (for example, \texttt{car insurance} and \texttt{travel insurance}), and an LLM turned each keyword into one question with a fixed instruction, for example \textit{``Quale assicurazione RC moto scegliere?''} (``Which motorbike liability insurance should I choose?''). 
Each prompt is asked without persona, as a baseline, and preceded by each of three buyer personas of its sector. 
A persona consists of three sentences describing life stage, experience with the category, and attitude to price. 
The Baseline persona (A) wants to spend as little as possible; the Price variant (B) changes only the price sentence, and the Profile variant (C) changes life stage or experience while keeping the same price sentence of A.

They should be considered as controlled purchase contexts, not demographic segments. 
Appendix Tables~\ref{tab:a_prompts} and~\ref{tab:a_personas} list prompts and personas.


We ask 20 prompts of each sector without a persona and with each of the three personas, with \texttt{search-off} and \texttt{search-on}, 10 times, to each of 3 engines with web search, \texttt{gpt-5.5}, \texttt{gemini-3.6-flash} and \texttt{perplexity/sonar} (Table~\ref{tab:engines}). 
This gives $80 \times 4 \times 2 \times 10 \times 3 = 19{,}200$ answers.
Additional details are in Appendix~\ref{app:collection}. 

\begin{table}[t]
\centering
\small
\caption{Engines of the two studies. \emph{Study~1}, \emph{Study~2}: whether the engine is used in the study; Study~1 runs each engine with \texttt{search-off} and \texttt{search-on}, Study~2 only with \texttt{search-off}. 
\emph{Ordered search actions}: whether the API returns the engine's successive searches in the order they were made. \emph{Searches}: share of the Study~1 calls with \texttt{search-on} in which the engine searched the web. In the results, we refer to the models as GPT, Gemini, Perplexity, and DeepSeek.}

\label{tab:engines}
\begin{tabular}{l c c c c}
\toprule
\textbf{Model} & \textbf{Study 1} & \textbf{Study 2} & \begin{tabular}[b]{@{}c@{}}\textbf{Ordered search actions}\end{tabular} & \textbf{Searches} \\
\midrule
\texttt{gpt-5.5} & \cmark & \cmark & \cmark & 98\% \\
\texttt{gemini-3.6-flash} & \cmark & \cmark & \xmark & 44\% \\
\texttt{perplexity/sonar} & \cmark & \cmark & \xmark & 100\% \\
\texttt{deepseek-v3.2} & \xmark & \cmark & -- & -- \\
\bottomrule
\end{tabular}
\end{table}

\spara{Metrics.}
\label{sec:m_s1_metrics}
%
A brand is \emph{mentioned} when it appears in the text of an answer, and \emph{cited} when its official site is among the answer's sources.
We define \emph{mention rate} of brand $k$, the proportion of answers that mention $k$ in the text at least once.
We define \emph{citation rate} of $k$, as the proportion of answers that cite the official website of $k$, limiting on the brands for which we know the official domain (683 out of 1,176 brands).
To account for the different prevalence of mentions and citations, we also normalize each rate by its average rate within the same sector and engine.
We compute brand's popularity by $\log_{10}(1+v)$, where $v$ is the mean monthly number of Google searches for its name in Italy over twelve months, obtained from DataForSEO\footnote{\url{https://dataforseo.com}, a commercial provider of search engine data.}.

We measure brand popularity bias as the slope of the mention rate (or citation rate) on popularity, estimated by Ordinary Least Squares (OLS) with one point per brand. If popularity bias equals 1, an increase of Google searches by 10 times implies an increase of 1 percentage point in the mention rate, i.e. one more answer out of 100.
The \emph{effect of Web search} on brand $k$ is its mention rate with \texttt{search-on} minus its mention rate with \texttt{search-off}. 

In Study~1, we also estimate how citation, mentions, and text vary with the same prompts and different personas.
We measure \textit{query similarity} as symmetric best-match cosine between their search queries, embedded with BGE-M3 \citep{chen2024bgem3}, a multilingual model, since the search queries are both in Italian and English. Two answers can issue a different number of search queries, so each search query is matched with the most similar one of the other answer, in both directions.
On the answer side, we measure the similarity between the set of retrieved brands (or sources) with Jaccard index.
Answer text similarity is measured with cosine similarity between their embeddings.


With \texttt{search-on}, we can obtain both mentions and citations.
A brand that is mentioned is \textit{found} if its name appears in the text of at least one returned source, it has \textit{no match} if it does not appear in sources and sources are available, and it \textit{cannot be verified} if at least one source cannot be read.

When using GPT, we can also inspect the order of actions deployed by the engine.
In particular, with \texttt{search-on}, we measure how often a brand appears in the query of first search.
Since GPT writes its first search query before receiving any web result, this behavior indicates that the engine drives the search biased by model's own knowledge, favouring the brand that appears in the first query.



A separate model (\texttt{gemini-2.5-flash}, temperature 0) lists the brands mentioned in each answer, and we merge by hand the spelling and product-line variants of the same brand; a manual check on 30 answers found 96\% of the extracted names correct (Appendix~\ref{app:collection}). 
A brand enters the analysis of a sector if at least one engine mentions it 5 or more times. 
We then compute its rates for all three engines. 
This gives 1,176 brands, 1,159 of them with a popularity value. 
As a robustness check, we repeat the analyses with a separate set of brands for each engine and with a threshold of 10 (see Appendix Table~\ref{tab:a_universe}). 
The source check uses the mentions of brands that reach the threshold in the engine itself.

In our analyses, the prompt is the sampled unit. 
Intervals are 95\% intervals from 1,000 bootstrap draws of the prompts within each sector, each draw keeping all the answers to the drawn prompts. Differences use the same draws. 
We compare the slopes with \texttt{search-off} and \texttt{search-on} by resampling brands within each configuration instead (2,000 draws; Appendix Table~\ref{tab:a_search_slopes}). 
Correlations across brands come with their standard $p$-values. 
We correct families of tests with Benjamini--Hochberg \citep{benjamini1995fdr}.

\subsection{Study 2: the controlled experiment}
\label{sec:m_s2}

In Study~2 we address RQ2 in a controlled experiment. An audit cannot tell whether an engine prefers a brand for its name or for its content, so here we fix the pages and change, on one of them, the name that signs it and the way it is written.
We include 5 product pages in the prompt, close to the mean number of sources per answer in Study~1 (on average 4.6 for GPT and 5.3 for Gemini; Perplexity returns all the results it retrieves, 9.0 on average), and ask the LLMs to rank them (with \texttt{search-off}), with different interventions.

\spara{Outcome.}
\label{sec:m_s2_outcome}
In each call a model receives a question and five pages in a given order, answers the question citing the pages and ends with a ranking of all five (Appendix~\ref{app:s2} gives the full message). 
One of the five is the \emph{target}, the page we change. 
An intervention effect is the number of positions gained by the target with the intervention. 
For example, if the target is third under its real brand and fourth under an invented brand, the effect is $-1$. 


We took a subsample of 24 prompts of Study~1, six per sector (Appendix Table~\ref{tab:a_s2}). 
These prompts ask the user to choose a product and have at least seven candidate brands. Brands that are mentioned in at least 5 of the 240 Study~1 answers to the prompt, with a verified official site, that sell the product under their own name. 
Among the most mentioned candidates, the most popular is set aside as the \emph{Better known} signature (see below), so it never supplies a page. The five most mentioned of the remaining candidates supply one product page each from their official site.

We use the full text of each page, without menus, reviews or related products. 
Before the experiment, a calibration run asked the four models to rank the five real pages.
The target is the page whose mean rank in this run is closest to 3, so that it has room to move both up and down. The other four pages never change.


We change the target in two ways. 
The \emph{signature} is the set of names on the page: the brand, the product and its technologies, codes and web address. 
The \emph{text} is the way the same facts are written. 
To intervene on the signature, we turn the text of the target into a template in which every name is a placeholder, and we fill the placeholders in five ways. 
\emph{Original brand Real product} uses the true brand and product, and \emph{Original brand Invented product} the true brand with an invented product. 
\emph{Better known Invented product} and \emph{Less known Invented product} use, respectively, a more popular and a less popular real brand, with the same invented product. \emph{Invented brand} uses an invented brand, whose \texttt{.it} domain does not exist, with the invented product. The five versions are identical word for word except for the names.


\emph{Better known} and \emph{Less known} brands are, respectively, the most popular among the brands most mentioned for the prompt, and the least popular brand mentioned for it; the target must be at least twice as popular as the less-known brand and at most half as popular as the better-known one. 
\emph{Invented brand} names share no distinctive part with the real ones. 
Appendix~\ref{app:s2} gives the full rules.


In \emph{Better known Invented product} and \emph{Less known Invented product}, a real brand with an invented product may itself look odd to a model. 
So, when the \emph{Better known} or \emph{Less known} brand sells a product of the same category as the target under a name of its own, we add \emph{Better known Real product} and \emph{Less known Real product}, which sign the page with that brand and that real product.
This is possible for both \emph{Better known} and \emph{Less known} brands in 15 of the 24 questions (Appendix Table~\ref{tab:a_s2_arms}). 
The product is the one of that brand most mentioned in the Study~1 answers to the prompt or, if there is none, the first suitable one on its official category page. 
In the other questions the target page mentions no product, the brand sells none with a name of its own, or, in travel, the product is a holiday tied to a place.


To intervene on the text, three rewrites keep every fact of the original and change only how it is written. 
\textsc{Structure} puts the decisive information first and uses headings and lists, while \textsc{Fluency} and \textsc{Authoritative} follow two methods of \citet{aggarwal2024geo}. 
\texttt{gpt-5.5} writes the rewrites on the template, so every signature carries the same rewrite. 
We keep a rewrite only if the placeholders and numbers are unchanged, its length is within 15\% of the original, so that a rewrite cannot gain or lose rank by adding or cutting text, and a different model (\texttt{deepseek-v3.2}) finds no fact added or removed (Appendix~\ref{app:s2}).


The placeholders remove the names, but the text may still reveal the brand, for example through a typical formula or a distinctive feature. 
To measure how often this happens, each model reads the target page with all names replaced, and we ask it which real brand the page belongs to (24 questions, 4 models, 3 repetitions). 
The question invites a guess that the ranking task does not, so the share of correct answers is an upper bound, as in the leak test of \citet{chen2026koudas}. 
In the rankings, we also count how often the answer mentions the original brand of the target when the page carries another signature.


For each question, we collect 5 signatures, with 4 texts, with 10 balanced orders of the five pages, each repeated twice, on 4 models. 
In addition to the three engines of Study~1 with \texttt{search-off}, we include the open-weight \texttt{deepseek-v3.2} (Table~\ref{tab:engines}).
We balance the orders so that each page appears in each position equally often. 
This gives $24 \times 5 \times 4 \times 20 \times 4 = 38{,}400$ rankings. 
The real-product versions add 9,600 rankings with the same texts, orders, repetitions and models.

Each effect is a difference in rank between two versions of the target in calls that are otherwise identical. 
We average these differences in three steps, first within each model and question, then over the four models, then over the 24 questions, so that every model and every question has the same weight. 
Because every signature is collected with every text, an effect of the signature is averaged over the four texts, and an effect of the text over the five signatures. 
This average summarises each factor fairly only if the two do not interact, so we also estimate the effect of rewriting under each signature (Figure~\ref{fig:r2}b and Appendix~\ref{app:s2_results}). 
The 24 questions are the sampled unit: we resample them 2,000 times to obtain 95\% intervals, and we test each hypothesis with a one-sided $t$ test on the 24 question means, adjusting the $p$-values with the Holm method~\citep{holm1979simple}.
\section{Results}
\label{sec:results}

\subsection{Study 1: the audit}
\label{sec:r1}

\begin{figure}[t]
  \centering
  \includegraphics[width=0.95\linewidth]{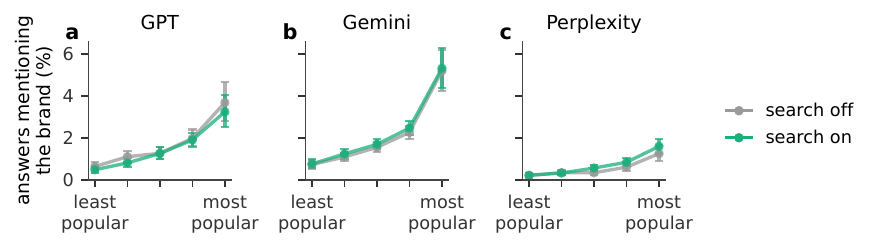}
  \Description{Three line charts, one per engine, of the share of answers mentioning a brand by quintile of popularity, with almost identical lines for \texttt{search-off} and \texttt{search-on}.}
  \caption{Mention rate as a function of popularity, with \texttt{search-off} and \texttt{search-on}, all sectors and personas together. 
  Share of answers that mention the brand, with \texttt{search-off} (grey) and \texttt{search-on} (green), by quintile of popularity within each engine and sector. Points are means over the brands of each quintile (about 240), with 95\% intervals from 1,000 prompt bootstrap samples, stratified by sector.}
  \label{fig:search}
\end{figure}
\subspara{H1: Generative search engines are affected by brand popularity bias.}
\label{sec:r1_h1}

\noindent If generative search engines are affected by brand popularity bias, the correlation between brand popularity and mention share should be positive (\HH{1a}), and so should the correlation between brand popularity and citation share (\HH{1b}). With \texttt{search-off} the engine can rely only on what it learned in training, so we also expect the correlation with mention share to be stronger with \texttt{search-off} than with \texttt{search-on} (\HH{1c}).

\begin{figure}[t]
  \centering
  \includegraphics[width=0.95\linewidth]{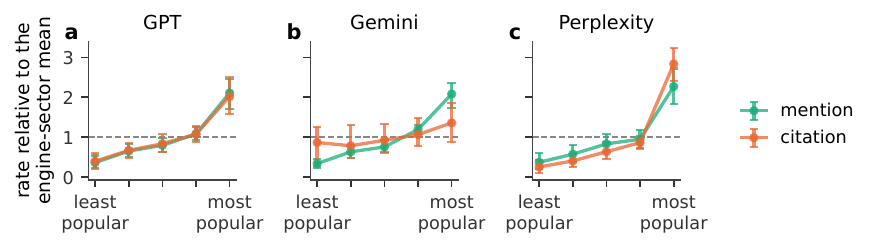}
  \Description{Three line charts, one per engine, of mention and citation rates relative to their engine-sector mean, by quintile of popularity, with a dashed line at the average brand. Both rates rise with popularity on all three engines, most in the top quintile.}
  \caption{Mention share and citation share by quintile of popularity, \texttt{search-on}, only for brands with a known domain. Each rate is divided by its mean over the brands of the same engine and sector.
  As such, 1 is the average brand (dashed line) and the levels of mention and citation are not comparable on this scale. Quintiles are formed within each engine and sector; points are means pooled over the four sectors, with 95\% intervals from 1,000 prompt bootstrap samples, stratified by sector.}
  \label{fig:mention_citation}
\end{figure}

\noindent Figure~\ref{fig:search} shows that the mention share rises with  brand popularity, from below 1\% in the least popular quintile to 1.2--5.3\% in the most popular, depending on the engine, with \texttt{search-off} and \texttt{search-on}. 
The correlation between popularity and mention rate is positive and significant in 23 of the 24 configurations (Spearman's $\rho$ up to 0.47, Benjamini--Hochberg), indicating that, among the brands the engines mention, more popular brands are mentioned more often. 
The exception is Perplexity in \texttt{Travel} with \texttt{search-off} ($\rho = 0.08$; its curve is in Appendix Figure~\ref{fig:a_search}).
\textbf{\HH{1a} is supported.} 
Figure~\ref{fig:mention_citation} puts mention and citation on the same relative scale. 
On all three engines both rise with popularity, and a brand in the most popular quintile is mentioned and cited 1.4 to 2.8 times as often as the average brand of its engine and sector. 
Relative to their own averages, mention and citation rise at the same rate on GPT, mention rises faster on Gemini and citation faster on Perplexity (Appendix Table~\ref{tab:a_norm}). 
The correlation between popularity and citation is significant in 8 of the 12 engine--sector pairs after Benjamini--Hochberg (Appendix Figure~\ref{fig:a_fits}). 
\textbf{\HH{1b} is partially supported}: it holds in these eight pairs, and not in \texttt{Sport} and \texttt{Travel} on GPT and on Gemini. 
In Appendix Figure~\ref{fig:a_mention_citation_raw}, we also show that mention and citation shares present wide variations across engines in absolute terms.
In Figure~\ref{fig:search} the curves with \texttt{search-off} and \texttt{search-on} overlap.
A brand searched ten times more often on Google is mentioned in 0.87 more answers out of 100 with \texttt{search-off} and in 0.89 more with \texttt{search-on} (difference +0.02 [$-$0.03, 0.07]; Appendix Figure~\ref{fig:a_search_scatter}).
\textbf{\HH{1c} is not supported.}

%

\subspara{H2: Search queries shape the sources of an answer, but not the brands it mentions.}
\label{sec:r1_route}

\noindent To see where the mentioned brands come from, we compare pairs of answers to the same prompt. If \HH{2} holds, answers whose search queries are more similar should also be more similar in their sources and text (\HH{2a}), while the similarity of the brands they mention should not grow with query similarity, since brands come mostly from what the model already knows (\HH{2b}).

\begin{figure}[t]
  \centering
  \includegraphics[width=0.95\linewidth]{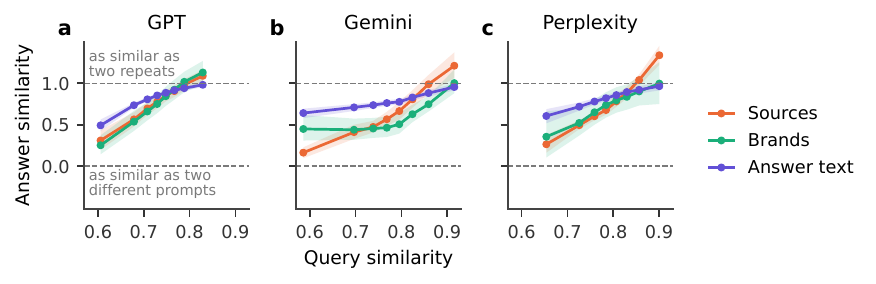}
  \Description{Three line charts, one per engine, of the similarity of sources, brands and answer text against the similarity of the search queries of two answers. All three similarities rise with query similarity, sources and brands most steeply.}
  \caption{Similarity of answers---in terms of sources, brands, and text---as a function of similarity of search queries. Query similarity matches each search query of one answer with the most similar search query of the other, and averages their cosine in both directions. Answer similarity is 
  rescaled so that 0 is the mean similarity of two answers to different prompts and 1 the mean similarity of two repeats (dashed lines). Values above 1 mean that two answers are more alike than two repeats on average, as happens when their search queries are very similar. Points: mean of the pairs in each of eight bands of query similarity; bands: 95\% intervals from resampling prompts.}
  \label{fig:h2}
\end{figure}

\begin{figure}[t]
  \centering
  \includegraphics[width=0.95\linewidth]{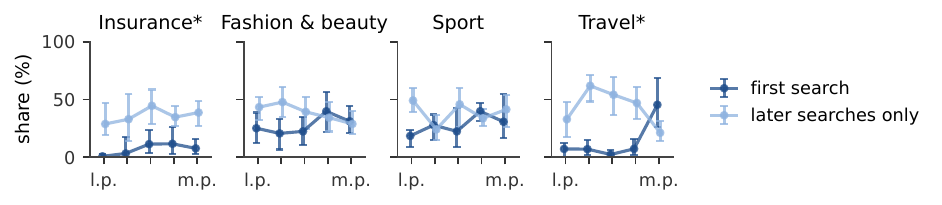}
  \Description{Four line charts, one per sector, of the share of brands mentioned by GPT whose name is in its first search action or only in later actions, by quintile of popularity.}
  \caption{GPT, \texttt{search-on}. For each quintile of popularity, the share of mentioned brands whose name is already in GPT's first search action (dark), or appears only in later actions (light). 
  Brands are split into quintiles of popularity within each sector, from least popular (l.p.) to most popular (m.p.). Bars: 95\% intervals from resampling prompts. $^{*}$ The first-action share of the most popular quintile differs from that of the least popular quintile, significant after Benjamini--Hochberg over the four sectors.}
  \label{fig:a_first_query}
\end{figure}

\noindent Figure~\ref{fig:h2} relates the similarity of two answers to the same prompt to the similarity of their search queries. 
On all three engines, the more similar the search queries, the more similar the sources, the mentioned brands and the text, and the slope is largest for sources, then brands, then text (on GPT 4.10, 3.57 and 2.30, on a scale where 0 means as alike as answers to different prompts and 1 as alike as two repeats). 
The relation holds also when we remove brand names from the search queries before measuring their similarity (Appendix~\ref{app:h2}).
\textbf{\HH{2a} is supported.} 
When two answers have more similar search queries, the overlap between their mentioned brands  also increase, almost as much as their sources (on GPT, slope 3.57 for brands against 4.10 for sources). This is hard to reconcile with brands coming only from the model's memory. \textbf{\HH{2b} is not supported.}
Two further observations, each on a different subset of calls, show where the names of the mentioned brands appear. 
In \texttt{Travel}, the share of mentioned brands already in GPT's first search action rises from 7\% in the least popular quintile to 45\% in the most popular (+0.38 [0.19, 0.63]); in \texttt{Insurance} the increase is smaller but also significant, from 1\% to 8\% (+0.07 [0.02, 0.15]) (Figure~\ref{fig:a_first_query}). 
Furthermore, when the engines search, the name of a mentioned brand appears in at least one returned source for 95.5\% of the mentions on GPT, 60.9\% on Gemini and 97.3\% on Perplexity; on Gemini, 24.9\% of the mentions could not be verified.
Notice that in 36.2\% of all checked mentions the name appears only on third-party pages, so a brand can be visible in the answer without being cited (Appendix~\ref{app:h2}).
%
%
Together with a brand popularity bias that is as strong with \texttt{search-on} as with \texttt{search-off}, these observations suggest that popular brands can reach the answer through the model's memory even with \texttt{search-on}.

%
\subspara{H3: The user persona does not affect the recommendations.}

\label{sec:r1_h3}
\noindent We then place a buyer persona before the question. If the persona does not affect the recommendations, the brand popularity bias should be the same with and without a persona context (\HH{3a}), and two answers to the same prompt with different personas should be as similar as two answers with the same persona (\HH{3b}).

\begin{figure}[t]
  \centering
  \includegraphics[width=0.95\linewidth]{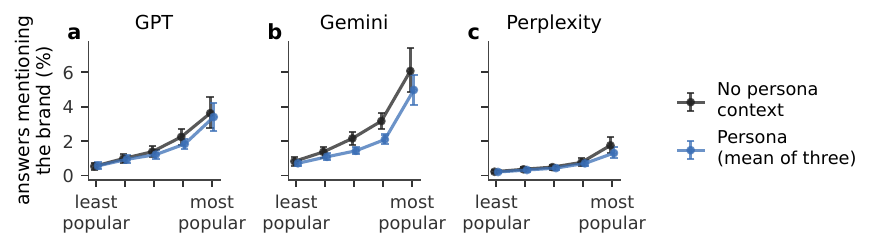}
  \Description{Three line charts, one per engine, of the mention rate by quintile of popularity without a persona context and averaged over the three personas. Both lines rise with popularity; the persona line is lower in the most popular quintiles.}
  \caption{Mention rate by quintile of popularity, without a persona context (black) and averaged equally over the three personas (blue), \texttt{search-off} and \texttt{search-on} together. Points: mean share of answers mentioning a brand, pooled over sectors; bars: 95\% intervals from 1,000 prompt bootstrap samples, stratified by sector.}
  \label{fig:persona}
\end{figure}

\noindent Figure~\ref{fig:persona} compares the mention rate by quintile of popularity without a persona context and with a persona. 
With a persona the curve is lower in the most popular quintiles, so the persona might reduce the brand popularity bias.
The slope of mention share on popularity falls from 0.98 without a persona context to 0.80 with the Baseline persona (A) and 0.85 with the Price variant (B), both significantly lower; with the Profile variant (C) it is 0.89, not significantly different (Appendix~\ref{app:h3}). 
Engine by engine and sector by sector, after the Benjamini--Hochberg correction only 12 of the 36 differences from no persona are significant, and in 11 of them the slope is lower with the persona (Appendix~\ref{app:h3}). 
\textbf{\HH{3a} is not supported} for the Baseline persona (A) and the Price variant (B).
Checking the search queries, we note that with the Baseline persona (A) or the Profile variant (C), at least one search query contains a price word in 86--96\% of the calls in \texttt{Fashion\&Beauty}, \texttt{Sport} and \texttt{Travel}, against 25--46\% without a persona context.
Two answers to the same prompt with different personas share fewer sources than two repeats with the same persona: their source similarity (Jaccard index) is lower by 0.07--0.08 on the three engines. 
Adjusting for the similarity of the search queries reduces the gap to 0.01--0.035 but does not close it (all intervals below zero). 
\textbf{\HH{3b} is not supported.}

\subsection{Study 2: the controlled experiment}
\label{sec:r2}


\begin{figure}[t]
  \centering
  \includegraphics[width=0.85\linewidth]{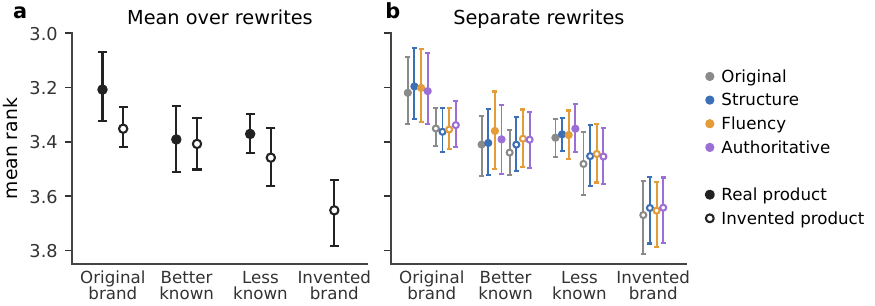}
  \Description{Two dot plots of the mean rank of the target page under each signature, with filled points for the real product name and hollow points for an invented one. Left: one point per signature, averaged over the four rewrites, from the true brand with its real product, ranked best, to the invented brand, ranked worst. Right: one point per signature and text; the four rewrites have almost identical ranks under every signature.}
  \caption{Study~2: mean rank of the target page (1 = ranked first; better ranks are higher in the plot) under each signature. 
  The horizontal axis represents the manipulation on the brand that signs the page.  
  Point style represents the manipulation on the product, filled points have a Real product name of the corresponding brand, and hollow points an Invented product name. (a) Mean over the original text and the three rewrites. (b) Colors encode the text rewriting strategy, including \emph{Original}, \emph{Structure}, \emph{Fluency}, and \emph{Authoritative}. Points: mean rank over the four models and over the questions in which each version exists (15 of the 24 for \emph{Better known Real product} and \emph{Less known Real product}). Bars: 95\% cluster bootstrap intervals over questions.}
  \label{fig:r2}
\end{figure}

\begin{table}[t]
\centering
\small
\caption{Study~2: change in the rank of the target page ($\Delta$, positions on a scale of 1--5, positive = ranked higher) between the two versions each row compares, named as in Figure~\ref{fig:r2} (\realdot{} Real product, \invdot{} Invented product). Each signature row is the difference between two points of Figure~\ref{fig:r2}a. Mean over the 24 questions of the mean over the four models; one-sided $p$, Holm-adjusted (n.s.: not significant). Effects of the signature average the four rewrites; effects of rewriting average the five signatures.}
\label{tab:r2}
\begin{tabularx}{\linewidth}{X l c c}
\toprule
\textbf{What changes on the target page} & & $\boldsymbol{\Delta}$ & \textbf{Holm} $p$ \\
\midrule
\multicolumn{4}{l}{\emph{Who signs it (Figure~\ref{fig:r2}a)}} \\
Original brand Real product \realdot{} vs Invented brand \invdot & \HH{4a} & \textbf{$+0.44$} & $<.01$ \\
Original brand Real product \realdot{} vs Original brand Invented product \invdot & \HH{4b} & \textbf{$+0.14$} & $<.05$ \\
Better known Invented product \invdot{} vs Less known Invented product \invdot & \HH{4c} & $+0.05$ & n.s. \\
Original brand Invented product \invdot{} vs mean of Better known and Less known Invented product \invdot & \HH{4d} & $+0.08$ & n.s. \\
Less known Invented product \invdot{} vs Invented brand \invdot & \HH{4e} & \textbf{$+0.19$} & $<.01$ \\
\midrule
\multicolumn{4}{l}{\emph{How it is written (Figure~\ref{fig:r2}b)}} \\
Rewritten vs original text (mean of the three GEO rewrites) & \HH{5a} & $+0.02$ & n.s. \\
$\Delta$ of \HH{4a} minus $\Delta$ of \HH{5a} & \HH{5b} & \textbf{$+0.42$} & $<.01$ \\
\bottomrule
\end{tabularx}
\end{table}

\subspara{H4: Generative search engines are affected by brand familiarity bias.}
\label{sec:r2_h4}

\noindent We compare versions of the same target page that differ in the names they carry (Section~\ref{sec:m_s2}; Figure~\ref{fig:r2}a).
To verify that our text manipulation cleanly neutralized brand identity, we tasked the model with recovering the original brand from the modified page text. 
The model guessed correctly in 38\% of cases; however, this recognition rate exhibits no relationship with the observed true-brand effect (Appendix~\ref{app:s2_results}).
If generative search engines are affected by brand familiarity bias, the page should rank higher with its true brand and product (\emph{Original brand Real product}) than with an invented brand and product (\emph{Invented brand}) (\HH{4a}). We then look for where this advantage comes from. With its true brand, the page should rank higher with its real product name than with an invented one (\emph{Original brand Real product} against \emph{Original brand Invented product}, \HH{4b}). When another real brand signs the page, with the same invented product, the text describes a product of a different brand. Neither of the two other real brands owns the page, so comparing them with each other cancels this mismatch. If popularity counts, the page should rank higher under the most popular (\emph{Better known Invented product} against \emph{Less known Invented product}, \HH{4c}). Their shared drop relative to the true brand with the same invented product (\emph{Original brand Invented product}) measures the mismatch itself: the page may rank lower because its text does not match the brand that signs it (\HH{4d}). Finally, with the same invented product, even a less popular real brand should rank the page higher than an invented brand (\emph{Less known Invented product} against \emph{Invented brand}, \HH{4e}).

\noindent Figure~\ref{fig:r2}a shows the mean rank of the target under each signature, averaged over the four rewrites: the target ranks best under its original brand and worst under an invented brand.
Averaged over questions, models and the four rewrites (Table~\ref{tab:r2}), the target page ranks 3.21 under its true brand and 3.65 under an invented one, so the true brand gains 0.44 positions [0.23, 0.69]. 
\textbf{\HH{4a} is supported.} 
The gain is positive for each of the four models, in each of the four sectors (Appendix Table~\ref{tab:a_s2_breakdown}).
%
%
%
%
Keeping the true brand but inventing the product name costs the page 0.14 positions [0.05, 0.27]. \textbf{\HH{4b} is supported.}
The rest of the gain of \HH{4a} comes from the brand name alone: with the same invented product, \emph{Original brand Invented product} ranks the page 0.30 positions above \emph{Invented brand} [0.13, 0.51] (Appendix Table~\ref{tab:a_s2_pairs}).
When another real brand signs the page, with the same invented product, the page drops compared with its true brand: by 0.06 positions under the more popular brand and by 0.11 under the less popular one.
\textbf{\HH{4c} is not supported.} 
\HH{4c} is the counterpart, at equal content, of the brand popularity bias of Study~1 (Section~\ref{sec:r1_h1}): among real brands that do not own the page, the more popular one does not rank it higher.
The drop the two brands share (0.08 [$-$0.02, 0.20]) is the cost of a brand that does not match the page (\HH{4d}), and it is not significant after the correction; this control also includes any recognition of the text as the original brand's.
A less popular real brand still ranks the page 0.19 positions above an invented brand [0.11, 0.30]. 
\textbf{\HH{4e} is supported.} 
\emph{Better known Real product} and \emph{Less known Real product} give the same picture: both brands rank the page above the invented brand (+0.27 and +0.29, significant after correction), not significantly below its true brand, and not differently from each other (Appendix Table~\ref{tab:a_s2_pairs}).

\subspara{H5: GEO rewriting increases the visibility of a page.}
\label{sec:r2_h6}

\noindent If GEO rewriting increases the visibility of a page, a rewritten page should rank higher than the same page with its original text (\HH{5a}). We also compare the two interventions on the same scale: changing brand name should raise the rank of the page more than GEO rewriting does (\HH{5b}).

Figure~\ref{fig:r2}b shows the mean rank of the target with each rewrite, under every signature. 
Each of the three rewrites moves the page by about 0.02 positions, and their average is +0.022 [$-$0.001, 0.049]. 
\textbf{\HH{5a} is not supported}, and the interval rules out gains larger than 0.05 positions. 
The original brand gains 0.42 positions more than rewriting [0.22, 0.66]. \textbf{\HH{5b} is supported.}
The rewrites, written by \texttt{gpt-5.5}, do not help more on GPT than on the other models (difference +0.005 [$-$0.03, 0.04]; Appendix Table~\ref{tab:a_s2_breakdown}).
 
\section{Discussion}
\label{sec:discussion}

In this work, we proposed an empirical framework for provider fairness in generative search in product recommendation.
Appendix Table~\ref{tab:hyp_summary} summarizes the outcome of each audited hypothesis. 

First, regarding \textbf{RQ1}, we find persistent \textit{brand popularity bias} across almost all evaluated engines and market sectors. 
Generative search engines disproportionately recommend market-dominant brands, a bias already present in their parametric memory, as shown by \texttt{search-off}. 
Crucially, enabling Retrieval-Augmented Generation (\texttt{search-on}) does not mitigate this head-item favoritism. 
In fact, (i) popularity bias remains unchanged between \texttt{search-on} and \texttt{search-off} conditions, and (ii) retrieval mechanisms often introduce popular brands in search queries before fetching external pages.
Consequently, external retrieval does not counteract the model's parametric brand bias.
Buyer personas are the only factor in our audit that weakens the brand popularity bias. A plausible explanation is that a persona turns a generic request into a constrained one: the engine has to match brands to a stated budget and profile, rather than list the best-known options in the category. 
Part of this may happen through retrieval, since with \texttt{search-on} the persona changes the search queries, which often add price terms, and the sources the engine retrieves.

Second, regarding \textbf{RQ2}, our controlled experiments demonstrate a multi-faceted \textit{brand familiarity bias}. 
Holding product descriptions, specifications, and layout strictly constant, authentic real-world brands achieve significantly higher recommendation ranks than synthetic brand controls. 
Our findings suggest that generative search engines may penalize unfamiliar entities that lack pre-existing weight representations in training data, creating a steep algorithmic barrier to entry.

Third, our experiments do not observe significant ranking gains from Generative Engine Optimization (GEO) rewrites. 
While prior work suggests that authority-style marketing can overcome brand dominance~\citep{chu2026incumbent}, those manipulations incorporated unverified or fabricated clinical claims. 
When GEO transformations are strictly constrained to be fact-preserving, surface-level textual optimization appears insufficient to overcome built-in parametric brand preferences.

Taken together, these results highlight key structural socio-economic risks in generative search. 
By over-representing dominant market players, generative search engines could feed a loop reminiscent of the Matthew effect (``rich-get-richer'')~\citep{klimashevskaia2024survey, mansoury2020feedback}. 
Because generative search engines synthesize choices into a few top recommendations rather than a paginated list of URLs, long-tail and local competitors may face near-total erasure.
At a global scale, exposing millions of daily users to similar or even identical, mode-collapsed recommendations risks creating digital market oligopolies and suppressing consumer choice diversity.

This systemic bias introduces a fundamental issue of market contestability. 
The observed \textit{brand familiarity bias} directly mirrors the item cold-start problem in traditional recommender systems~\citep{schein2002coldstart}.
However, unlike traditional systems where explicit product features can bridge the cold-start gap, generative search engines rank an unknown name lower, as a novel entrant would have, even when complete product descriptions are provided in-context. 
Consequently, newly established brands or sustainable startups may be disadvantaged simply because they were founded after model pre-training.


\spara{Broader implications.}
\label{sec:d_implications}
%
%
%
%
%
Our experimental pipeline offers a low-cost audit that developers of generative search engines could run before deployment, comparing the rank variations across controlled identity swaps. 
The design is not specific to brands: in principle it applies to any entity whose name a model may know, such as news outlets, public bodies or authors, and prior work already finds that the name of a source can matter more than its content~\citep{dai2025media, khan2026agents}.
For product vendors and marketers, our findings set boundaries on the efficacy of Generative Engine Optimization (GEO). 
While marketing agencies advocate stylistic rewriting for LLM visibility, fact-preserving GEO yields marginal returns, and a new provider may gain more from becoming known in the web sources that models learn from and retrieve.
For regulators, focusing solely on citation transparency overlooks how visibility is allocated: 
auditing frameworks must therefore evaluate inline entity mentions alongside source links to safeguard market contestability and antitrust oversight.
Finally, consumers relying on generative search for product discovery may be exposed to a narrower, head-biased subset of the market. 
Awareness of these built-in parametric priors may help users seeking specialized, local or emerging alternatives.

\spara{Limitations.}
\label{sec:limitations}
Our empirical findings should be contextualized within several deliberate methodological trade-offs. 
First, outputs from black-box LLMs inherently fluctuate across time and geography due to continuous background updates and localized indexing~\citep{chen2026koudas, schulte2026measure}. 
To control for these intrinsic auditing dynamics, we fixed all prompts to a single temporal snapshot and location (Italian queries from Rome), leaving cross-cultural or regional brand disparities~\citep{betello2026passport} outside our scope. 
Second, in Study~2, we explicitly executed experiments with live web retrieval disabled (\texttt{search-off}). 
While this setup omits live search reranking in the wild, this choice allows us to exert total experimental control over the in-context page, thus isolating brand bias. 
Third, our persona conditioning relied on three holistic multi-sentence templates rather than an exhaustive factorial grid. 
While this captures only a subset of possible user profiles, the three personas lower the brand popularity bias in the same direction, significantly for two of them.
Fourth, model audits were conducted through vendor API endpoints. 
While APIs provide reproducible, programmatic access, they may omit proprietary, client-side system prompts or interface heuristics present in consumer-facing chat interfaces.
Fifth, our GEO rewrites strictly operationalized established optimization strategies~\citep{aggarwal2024geo} under an explicit fact-preserving constraint. 
We deliberately excluded deceptive or fabricated claims (e.g., fake statistics) to reflect legitimate on-page optimization, leaving whether adversarial copy can bypass parametric priors to future work.
Sixth, our evaluation focuses exclusively on generative search, precluding a direct empirical baseline against traditional search engines. 
Finally, we quantified brand visibility via mention and citation frequencies without performing sentiment analysis. 
While negative portrayals are theoretically plausible, manual inspection of the generated outputs revealed no instances of unfavorable brand mentions or citations; recommendations were uniformly positive or neutral.

\spara{Future Work.}
\label{sec:future}
Future research should extend this audit methodology along three main avenues. 
First, conducting \textit{relevance-conditional audits} would compare generated brand recommendations against an independent, ground-truth inventory of market offers matched for price and specification. 
Second, testing real-world brand entities founded strictly after the model's training cutoff date under \texttt{search-on} conditions would provide a sharper delineation between real-world brand emergence and synthetic entity penalties. Finally, expanding this framework to open-weight architectures and multi-agent retrieval pipelines could pinpoint the exact layer dynamics where parametric brand priors override retrieved in-context specifications.

\section*{Endmatter}
\subsection*{Ethical considerations and adverse impacts}
The studies query deployed commercial systems through their official APIs and involve no human participants and no personal data. Source pages were read after the calls, to check whether the names of mentioned brands appear in them (Study~1) and to build the product pages of Study~2. Each page was first requested with a declared user agent; pages that blocked this request were retrieved through a commercial unblocking service (Bright Data Web Unlocker) or from Internet Archive snapshots, and the few still unreadable were saved manually. Invented brand names were checked to have no existing or registered web domain. 
Showing what moves a page in a generated ranking has a dual use: it can inform providers and regulators, and it can guide manipulation; rewriting moves pages very little, but the identity result suggests that signing a page with a known brand works, which is impersonation and is not something we recommend or tested in the wild.

\subsection*{Generative AI usage statement}
The language models studied are the objects of the audit and of the experiment (Section~\ref{sec:methods}). \texttt{gemini-2.5-flash} extracted brand names from the answers; \texttt{gpt-5.5} extracted the text of the Study~2 pages (\texttt{gemini-3.6-flash} for one non-target page that \texttt{gpt-5.5} did not process) and wrote the rewrites, and \texttt{deepseek-v3.2} checked them for added or removed facts; LLMs also classified prompts and candidate brands during selection, with fixed instructions. We used Claude (Anthropic) as a coding assistant for the analysis scripts, and as a writing assistant for drafting and editing parts of the text. The authors designed the studies, checked the results against the data, and reviewed and take responsibility for all code and text.

\subsection*{Data and code availability}
Code, prompts, product pages, and aggregated data are available at \url{https://anonymous.4open.science/r/brand-bias-generative-search-3E74}.

\bibliographystyle{ACM-Reference-Format}
\bibliography{references}

\appendix
\counterwithin{figure}{section}
\counterwithin{table}{section}
\section{Study 1: prompts and personas}
\label{app:prompts}

\begingroup\scriptsize
\begin{longtable}{p{0.11\linewidth} p{0.47\linewidth} p{0.24\linewidth} r}
\caption{The 80 prompts of Study~1, with the keyword they were derived from and its mean monthly search volume in Italy over twelve months (DataForSEO).}\label{tab:a_prompts}\\
\toprule ID & Prompt & Keyword & Volume \\ \midrule \endfirsthead
\toprule ID & Prompt & Keyword & Volume \\ \midrule \endhead
\bottomrule \endlastfoot
assicurazioni\_01 & \textit{Quale assicurazione RC moto scegliere?} & assicurazione rc moto & 133 \\
assicurazioni\_02 & \textit{Quale assicurazione annullamento viaggio mi conviene scegliere in base a cosa copre?} & assicurazione annullamento viaggio cosa copre & 102 \\
assicurazioni\_03 & \textit{Quale assicurazione bici dovrei scegliere?} & assicurazione bici & 695 \\
assicurazioni\_04 & \textit{Qual è la migliore assicurazione casa con responsabilità civile da scegliere?} & assicurazione casa responsabilità civile & 249 \\
assicurazioni\_05 & \textit{Quale polizza famiglia mi consigliate?} & polizza famiglia & 104 \\
assicurazioni\_06 & \textit{Qual è la migliore assicurazione kasko da scegliere?} & migliore assicurazione kasko & 118 \\
assicurazioni\_07 & \textit{Qual è la migliore assicurazione mutuo da scegliere?} & assicurazione mutuo & 857 \\
assicurazioni\_08 & \textit{Qual è la migliore assicurazione mensile per moto che potrei scegliere?} & assicurazione mensile per moto & 1,972 \\
assicurazioni\_09 & \textit{Quale assicurazione sui voli mi consigli di scegliere?} & assicurazione sui voli & 702 \\
assicurazioni\_10 & \textit{Qual è la migliore assicurazione natante da scegliere?} & assicurazione natante & 308 \\
assicurazioni\_11 & \textit{Quali sono le migliori opzioni di assicurazione per la famiglia?} & assicurazione per la famiglia & 540 \\
assicurazioni\_12 & \textit{Quali sono i migliori preventivi di assicurazioni online?} & preventivi assicurazioni on line & 470 \\
assicurazioni\_13 & \textit{Qual è l'assicurazione auto migliore da scegliere?} & assicurare auto & 182 \\
assicurazioni\_14 & \textit{Qual è la migliore polizza sul mutuo da scegliere?} & polizza sul mutuo & 158 \\
assicurazioni\_15 & \textit{Qual è la migliore polizza cristalli senza RCA da scegliere?} & polizza cristalli senza rca & 226 \\
assicurazioni\_16 & \textit{Quale assicurazione è la migliore per i medici specializzandi?} & assicurazione medici specializzandi & 119 \\
assicurazioni\_17 & \textit{Qual è l'assicurazione RCA migliore da scegliere?} & assicurazione rca & 3,442 \\
assicurazioni\_18 & \textit{Quali sono le migliori opzioni di assicurazione personale infortuni da scegliere?} & assicurazione personale infortuni & 291 \\
assicurazioni\_19 & \textit{Quale assicurazione spese veterinarie dovrei scegliere?} & assicurazione spese veterinarie & 438 \\
assicurazioni\_20 & \textit{Quali sono le migliori opzioni per l'assicurazione vita in base ai costi?} & costi assicurazione vita & 812 \\
viaggi\_01 & \textit{Quale crociera nel Mediterraneo mi consigli di scegliere in base al prezzo?} & prezzo crociera mediterraneo & 787 \\
viaggi\_02 & \textit{Quali sono le migliori crociere a poco prezzo?} & crociere a poco prezzo & 134 \\
viaggi\_03 & \textit{Quali sono i migliori villaggi last minute nel Salento?} & villaggi salento last minute & 143 \\
viaggi\_04 & \textit{Quali villaggi scegliere per le vacanze in Cilento?} & vacanze in cilento villaggi & 305 \\
viaggi\_05 & \textit{Quali sono i migliori villaggi sul mare nel Conero da scegliere?} & villaggi conero sul mare & 585 \\
viaggi\_06 & \textit{Quali sono i migliori villaggi turistici sul mare nel Gargano?} & villaggi turistici gargano sul mare & 121 \\
viaggi\_07 & \textit{Quale villaggio turistico all inclusive mi consiglieresti?} & villaggio turistico all inclusive & 340 \\
viaggi\_08 & \textit{Quali sono i migliori villaggi turistici all inclusive in Riviera Romagnola?} & villaggi turistici riviera romagnola all inclusive & 157 \\
viaggi\_09 & \textit{Quali sono le migliori offerte di noleggio disponibili?} & offerte noleggio & 318 \\
viaggi\_10 & \textit{Quali sono i migliori tour operator per viaggi di gruppo organizzati?} & tour operator viaggi di gruppo organizzati & 162 \\
viaggi\_11 & \textit{Quali sono le migliori offerte auto a noleggio tra cui scegliere?} & offerte auto a noleggio & 106 \\
viaggi\_12 & \textit{Quali sono le migliori crociere con offerte scontatissime?} & crociere scontatissime & 306 \\
viaggi\_13 & \textit{Quali sono i migliori villaggi all inclusive alle Maldive del Salento?} & maldive del salento villaggi all inclusive & 222 \\
viaggi\_14 & \textit{Quali sono le migliori offerte per le crociere?} & offerte crociere & 8,125 \\
viaggi\_15 & \textit{Quali sono i migliori tour operator per un viaggio a Zanzibar?} & tour operator zanzibar & 378 \\
viaggi\_16 & \textit{Quali sono le migliori offerte per crociere di Capodanno?} & offerte crociere capodanno & 200 \\
viaggi\_17 & \textit{Quali sono le migliori opzioni per il noleggio auto con prezzi bassi?} & noleggio auto prezzi bassi & 191 \\
viaggi\_18 & \textit{Quali sono i migliori tour operator siciliani da scegliere per un viaggio?} & tour operator siciliani & 439 \\
viaggi\_19 & \textit{Quali sono i migliori voli disponibili, considerando i prezzi?} & prezzi voli & 1,003 \\
viaggi\_20 & \textit{Quali sono le migliori compagnie crociere italiane tra cui scegliere?} & compagnie crociere italiane & 162 \\
moda\_beauty\_01 & \textit{Quali sono le migliori scarpe eleganti comode con tacco medio da scegliere?} & scarpe eleganti comode con tacco medio & 133 \\
moda\_beauty\_02 & \textit{Qual è il contorno occhi antirughe migliore da scegliere?} & contorno occhi antirughe migliore & 271 \\
moda\_beauty\_03 & \textit{Quali profumi originali con prezzi bassissimi dovrei scegliere?} & profumi originali a prezzi bassissimi & 778 \\
moda\_beauty\_04 & \textit{Quale abbigliamento on line mi consigliate di acquistare?} & abbigliamento on line & 1,567 \\
moda\_beauty\_05 & \textit{Qual è il miglior siero antimacchie da scegliere?} & miglior siero antimacchie & 527 \\
moda\_beauty\_06 & \textit{Quali sono i migliori giubbotti in pelle da uomo che potrei scegliere?} & i migliori giubbotti in pelle uomo & 392 \\
moda\_beauty\_08 & \textit{Quali sono i migliori jeans momfit da scegliere?} & momfit & 464 \\
moda\_beauty\_09 & \textit{Quali vestiti di marca a poco prezzo dovrei scegliere?} & vestiti di marca a poco prezzo & 241 \\
moda\_beauty\_10 & \textit{Quale top verde elegante dovrei scegliere?} & top verde elegante & 118 \\
moda\_beauty\_11 & \textit{Quali maschere viso naturali dovrei scegliere?} & maschere viso naturali & 364 \\
moda\_beauty\_12 & \textit{Quali sono le migliori offerte di abbigliamento per uomo da scegliere?} & offerte abbigliamento uomo & 542 \\
moda\_beauty\_13 & \textit{Qual è la migliore maschera viso al collagene da scegliere?} & maschera viso collagene migliore & 332 \\
moda\_beauty\_14 & \textit{Qual è la migliore crema contorno occhi da scegliere?} & crema contorno occhi migliore & 2,342 \\
moda\_beauty\_15 & \textit{Quali sono le migliori borse in offerta o con sconti da scegliere?} & borse scontate & 1,170 \\
moda\_beauty\_16 & \textit{Quali sono le migliori pochette di marca da outlet tra cui scegliere?} & pochette firmate outlet & 427 \\
moda\_beauty\_18 & \textit{Qual è il retinolo migliore?} & retinolo migliore & 892 \\
moda\_beauty\_19 & \textit{Quali sono i migliori fondotinta da scegliere?} & migliori fondotinta & 355 \\
moda\_beauty\_21 & \textit{Quali sono i migliori jeans push up da scegliere?} & jeans push up migliori & 152 \\
moda\_beauty\_22 & \textit{Quali sono i migliori giubbotti invernali da uomo tra cui scegliere?} & migliori giubbotti invernali uomo & 366 \\
moda\_beauty\_23 & \textit{Quali sono i profumi buoni da uomo che mi consiglieresti?} & profumi buoni uomo & 690 \\
sport\_01 & \textit{Quali sono le migliori scarpe da trekking da scegliere?} & le migliori scarpe da trekking & 200 \\
sport\_02 & \textit{Quale offerta di bici elettrica dovrei scegliere?} & offerta bici elettrica & 636 \\
sport\_03 & \textit{Quali sono le differenze tra un tapis roulant elettrico e uno magnetico, e quale mi conviene scegliere?} & differenze tra tapis roulant elettrico e magnetico & 128 \\
sport\_04 & \textit{Quali sono le migliori scarpe da corsa su strada che dovrei scegliere?} & migliori scarpe da corsa su strada & 103 \\
sport\_05 & \textit{Quali sono le migliori scarpe da trail running per uomo?} & scarpe trail running uomo migliori & 209 \\
sport\_06 & \textit{Quali sono le migliori e-bike in base al costo?} & prezzi e bike & 161 \\
sport\_07 & \textit{Quali sono i migliori tapis roulant per casa da scegliere?} & tapis roulant migliori per casa & 178 \\
sport\_08 & \textit{Qual è la migliore scarpa running da scegliere?} & migliore scarpa running & 172 \\
sport\_09 & \textit{Qual è la migliore ebike per rapporto qualità prezzo?} & migliore ebike qualità prezzo & 218 \\
sport\_10 & \textit{Quale tapis roulant dovrei scegliere in base al prezzo?} & prezzo tapis roulant & 217 \\
sport\_11 & \textit{Quali sono i migliori tapis roulant da scegliere?} & tapis roulant i migliori & 605 \\
sport\_12 & \textit{Quali e-bike con offerte mi consiglieresti?} & e bike offerte & 346 \\
sport\_13 & \textit{Qual è la racchetta da padel migliore da scegliere?} & racchetta padel migliore & 157 \\
sport\_14 & \textit{Quale mountain bike da 24 pollici in offerta dovrei scegliere?} & mtb 24 pollici offerta & 132 \\
sport\_15 & \textit{Quale racchetta da tennis scegliere in base al prezzo?} & racchetta da tennis prezzi & 228 \\
sport\_16 & \textit{Quali sono i migliori outfit per la palestra tra cui scegliere?} & outfit palestra & 469 \\
sport\_17 & \textit{Quali sono le migliori cargo bike economiche da scegliere?} & cargo bike economiche & 288 \\
sport\_20 & \textit{Qual è la racchetta da tennis migliore da scegliere?} & racchetta tennis migliore & 608 \\
sport\_21 & \textit{Quali sono le migliori scarpe da corsa da scegliere?} & migliori scarpe da corsa & 630 \\
sport\_22 & \textit{Quali sono le migliori scarpe da calcio a poco prezzo?} & scarpe calcio poco prezzo & 160 \\
\end{longtable}\endgroup

\begin{table}[ht]
\centering\scriptsize
\caption{The twelve buyer personas of Study~1, placed before the question in Italian; English translation below each.}\label{tab:a_personas}
\begin{tabularx}{\linewidth}{l l X}
\toprule Sector & Persona & Text \\ \midrule
\texttt{Insurance} & Baseline persona (A) & \textit{Sono genitore con una famiglia a carico. È la prima volta che stipulo una polizza. Cerco di spendere il meno possibile.} \newline I am a parent with a family to support. It is the first time I take out a policy. I try to spend as little as possible. \\
\texttt{Insurance} & Price variant (B) & \textit{Sono genitore con una famiglia a carico. È la prima volta che stipulo una polizza. Spendo volentieri di più per la qualità.} \newline I am a parent with a family to support. It is the first time I take out a policy. I am happy to spend more for quality. \\
\texttt{Insurance} & Profile variant (C) & \textit{Sono al primo anno di patente. È la prima volta che stipulo una polizza. Cerco di spendere il meno possibile.} \newline I am in my first year with a driving licence. It is the first time I take out a policy. I try to spend as little as possible. \\
\texttt{Travel} & Baseline persona (A) & \textit{Sono single e viaggio senza figli. Ho un'esperienza nella media con i viaggi. Cerco di spendere il meno possibile.} \newline I am single and travel without children. I have average experience with travel. I try to spend as little as possible. \\
\texttt{Travel} & Price variant (B) & \textit{Sono single e viaggio senza figli. Ho un'esperienza nella media con i viaggi. Cerco il miglior rapporto tra qualità e prezzo.} \newline I am single and travel without children. I have average experience with travel. I look for the best value for money. \\
\texttt{Travel} & Profile variant (C) & \textit{Sono genitore di figli piccoli. Ho un'esperienza nella media con i viaggi. Cerco di spendere il meno possibile.} \newline I am a parent of young children. I have average experience with travel. I try to spend as little as possible. \\
\texttt{Fashion\&Beauty} & Baseline persona (A) & \textit{Sono genitore di figli piccoli. Ho un'esperienza nella media con moda e cosmetici. Cerco di spendere il meno possibile.} \newline I am a parent of young children. I have average experience with fashion and cosmetics. I try to spend as little as possible. \\
\texttt{Fashion\&Beauty} & Price variant (B) & \textit{Sono genitore di figli piccoli. Ho un'esperienza nella media con moda e cosmetici. Cerco il miglior rapporto tra qualità e prezzo.} \newline I am a parent of young children. I have average experience with fashion and cosmetics. I look for the best value for money. \\
\texttt{Fashion\&Beauty} & Profile variant (C) & \textit{Sono single e non ho figli. Ho un'esperienza nella media con moda e cosmetici. Cerco di spendere il meno possibile.} \newline I am single and have no children. I have average experience with fashion and cosmetics. I try to spend as little as possible. \\
Sport & Baseline persona (A) & \textit{Sono genitore di figli piccoli. Ho cominciato da poco a fare sport. Cerco di spendere il meno possibile.} \newline I am a parent of young children. I started doing sport only recently. I try to spend as little as possible. \\
Sport & Price variant (B) & \textit{Sono genitore di figli piccoli. Ho cominciato da poco a fare sport. Cerco il miglior rapporto tra qualità e prezzo.} \newline I am a parent of young children. I started doing sport only recently. I look for the best value for money. \\
Sport & Profile variant (C) & \textit{Sono genitore di figli piccoli. Faccio sport a un livello intermedio. Cerco di spendere il meno possibile.} \newline I am a parent of young children. I do sport at an intermediate level. I try to spend as little as possible. \\
\bottomrule
\end{tabularx}
\end{table}

\section{Data collection}
\label{app:collection}

All calls go through the official APIs, with no system prompt and the provider's default parameters; in Study~1 the user location is Rome, Italy.

All Study~1 calls were made on 14 October 2026 (UTC). In Study~2, the rankings were collected between 28 October and 3 November 2026, and the calibration run and the brand ratings between 23 and 25 October. The brand extractor (\texttt{gemini-2.5-flash}) was checked by hand on 30 answers from a pilot run: 151 of the 157 extracted names were correct brands (96\%), and it missed 2 of the 153 brands in those answers (1\%). In the source check, a page counts as readable if it is not an anti-bot check and its extracted text has at least 1,000 characters, which excludes pages whose content did not load. Every answer stores the model version returned by the API and, in Study~1, the search queries, the sources and the text. Prompts, personas, pages, message templates and instructions were fixed before the calls, and their hashes are stored with the answers.

\section{Study 1: additional results}
\label{app:figs}

\subsection{Brand popularity bias in mention and citation (H1)}
\label{app:h1}
\begin{figure}[ht]
  \centering
  \includegraphics{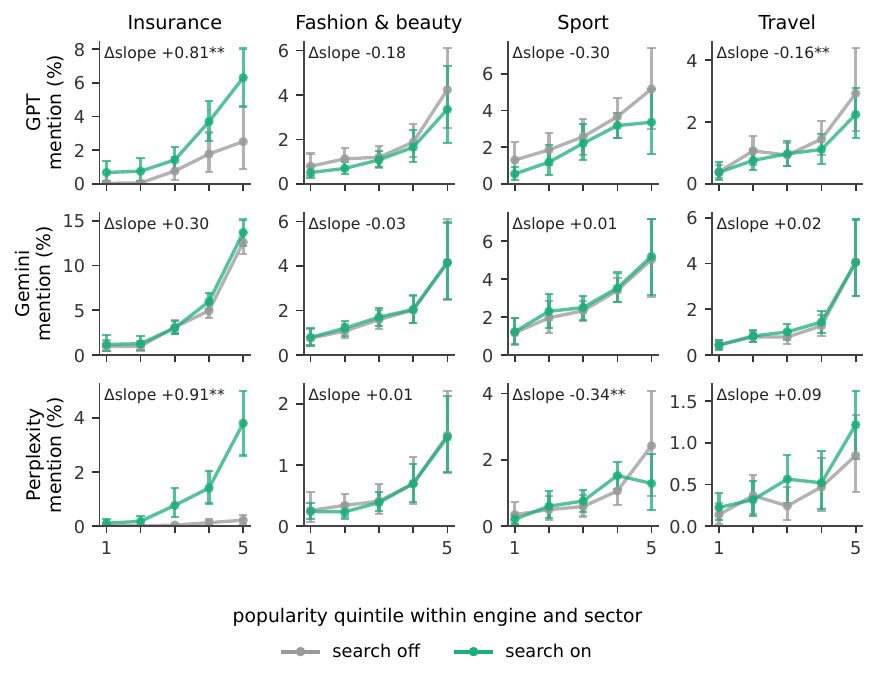}
  \Description{Grid of line charts, one per engine and sector, of the mention rate by quintile of popularity with search off and on.}
  \caption{Mention rate by quintile of popularity with search off (grey) and on (green), by engine and sector; 95\% intervals from 1,000 prompt bootstrap samples, stratified by sector. $\Delta$slope: difference between the two slopes of Table~\ref{tab:a_search_slopes}; $^{**}$ significant after Benjamini--Hochberg over the 12 engine--sector pairs, $^{*}$ $p < 0.05$ before correction. With search off, Perplexity almost never mentions an insurer.}
  \label{fig:a_search}
\end{figure}

By engine and sector, the slopes of mention on popularity with search off and on differ in 4 of the 12 pairs after Benjamini--Hochberg (Table~\ref{tab:a_search_slopes}, Figure~\ref{fig:a_search}). With search on, the slope is steeper for insurance on GPT (+0.81) and on Perplexity (+0.91), which with search off almost never mentions an insurer, and flatter for sport on Perplexity ($-$0.34) and for travel on GPT ($-$0.16). On Gemini, the calls that actually searched show a flatter slope than the calls with search off (0.95 against 1.34; difference $-$0.38, $p < 0.001$, OLS with errors clustered by brand), in all four sectors. Gemini chooses which calls search, so this comparison is not controlled.

\begin{figure}[ht]
  \centering
  \includegraphics{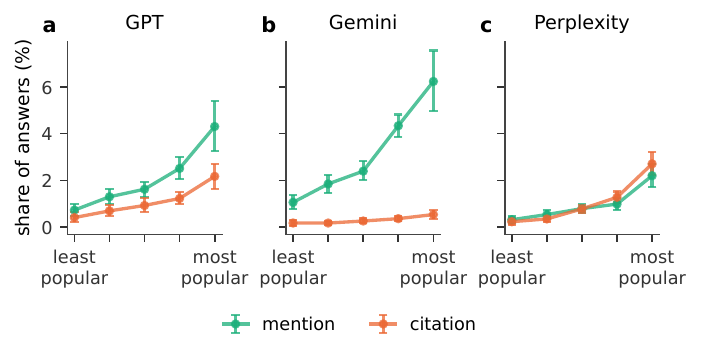}
  \Description{Three line charts, one per engine, of mention and citation rates in percent by quintile of popularity. On GPT both rise; on Gemini mention rises to about 6 percent while citation stays below 1 percent; on Perplexity the two lines overlap.}
  \caption{Mention and citation by quintile of popularity in absolute terms, search on, for brands with a known domain. Quintiles are formed within each engine and sector. Points are the mean share of answers that mention the brand (green) or cite it (orange), pooled over the four sectors, with 95\% intervals from 1,000 prompt bootstrap samples, stratified by sector. On Perplexity the sources are the retrieved results, so its citation share can exceed its mention share.}
  \label{fig:a_mention_citation_raw}
\end{figure}

Figure~\ref{fig:a_mention_citation_raw} shows the rates of Figure~\ref{fig:mention_citation} before normalisation. GPT mentions a brand about twice as often as it cites it, Gemini mentions brands often and cites them rarely, and on Perplexity, whose sources are the retrieved results, the citation rate can exceed the mention rate. A mentioned brand with a known domain is cited in 32--64\% of the answers that searched on GPT, 15--80\% on Perplexity and 3--28\% on Gemini, depending on the sector. Because the levels differ, the raw slopes of mention and citation on popularity are not directly comparable; Table~\ref{tab:a_norm} compares the slopes of the normalised rates. On GPT the two normalised slopes are the same (0.52 and 0.51), so the steeper raw slope of mention comes from its higher level. On Gemini popularity counts more for mention than for citation (difference +0.38 [0.18, 0.58]). On Perplexity it counts more for citation ($-$0.26 [$-$0.44, $-$0.10]).

\begin{table}[ht]
\centering\small
\caption{Study~1: popularity bias under different sets of brands. Slopes as in Table~\ref{tab:a_search_slopes}, all engines and sectors together. Main analysis: one set of brands per sector, made of the brands that at least one engine mentions in 5 or more of its 1,600 answers in that sector, with zeros for the engines that do not name them. Alternatives: a separate set for each engine (brands that this engine mentions at least 5 times), and a threshold of 10. A threshold below 5 is not possible, because popularity was measured only for the brands above it.}
\label{tab:a_universe}
\begin{tabular}{l r r r r l}
\toprule
Set of brands & Brand rows & Search off & Search on & Difference & 95\% interval \\
\midrule
Per sector, $\geq 5$ (main) & 3,567 & +0.87 & +0.89 & +0.02 & $[-0.03, 0.07]$ \\
Per engine, $\geq 5$ & 2,101 & +1.34 & +1.35 & +0.01 & $[-0.07, 0.10]$ \\
Per sector, $\geq 10$ & 2,463 & +1.05 & +1.08 & +0.03 & $[-0.04, 0.11]$ \\
Per engine, $\geq 10$ & 1,450 & +1.61 & +1.62 & +0.01 & $[-0.11, 0.12]$ \\
\bottomrule
\end{tabular}
\end{table}

\begin{table}[ht]
\centering\small
\caption{Study~1: how many percentage points the mention rate of a brand rises for each tenfold increase in its search volume, with search off and with search on. Each slope is an OLS fit with one point per brand; the difference between the two slopes is also the slope of the per-brand effect of search (Figure~\ref{fig:a_search_scatter}). Rows for all sectors and for all engines add engine $\times$ sector fixed effects. 95\% intervals from resampling brands within each engine and sector; $p$ from a $t$ test with robust (HC3) errors, corrected with Benjamini--Hochberg over the 12 engine--sector rows (summary rows uncorrected).}
\label{tab:a_search_slopes}
\begin{tabular}{l r r r l r}
\toprule
 & Search off & Search on & Difference & 95\% interval & $p$ \\
\midrule
\multicolumn{6}{l}{\emph{GPT}} \\
\quad Insurance & +0.71 & +1.52 & +0.81 & $[0.27, 1.49]$ & 0.031 \\
\quad Fashion and beauty & +1.09 & +0.91 & $-$0.18 & $[-0.36, 0.00]$ & 0.103 \\
\quad Sport & +1.41 & +1.10 & $-$0.30 & $[-0.61, 0.03]$ & 0.103 \\
\quad Travel & +0.70 & +0.54 & $-$0.16 & $[-0.28, -0.05]$ & 0.031 \\
\quad \emph{All sectors} & +0.94 & +0.86 & $-$0.07 & $[-0.19, 0.03]$ & 0.205 \\
\midrule
\multicolumn{6}{l}{\emph{Gemini}} \\
\quad Insurance & +2.95 & +3.25 & +0.30 & $[0.03, 0.62]$ & 0.103 \\
\quad Fashion and beauty & +1.11 & +1.08 & $-$0.03 & $[-0.13, 0.06]$ & 0.617 \\
\quad Sport & +1.38 & +1.38 & +0.01 & $[-0.14, 0.16]$ & 0.953 \\
\quad Travel & +1.04 & +1.06 & +0.02 & $[-0.04, 0.08]$ & 0.609 \\
\quad \emph{All sectors} & +1.34 & +1.37 & +0.04 & $[-0.02, 0.10]$ & 0.235 \\
\midrule
\multicolumn{6}{l}{\emph{Perplexity}} \\
\quad Insurance & +0.06 & +0.97 & +0.91 & $[0.41, 1.50]$ & 0.019 \\
\quad Fashion and beauty & +0.39 & +0.40 & +0.01 & $[-0.13, 0.16]$ & 0.936 \\
\quad Sport & +0.79 & +0.45 & $-$0.34 & $[-0.61, -0.10]$ & 0.031 \\
\quad Travel & +0.21 & +0.31 & +0.09 & $[-0.01, 0.23]$ & 0.222 \\
\quad \emph{All sectors} & +0.34 & +0.43 & +0.10 & $[0.00, 0.20]$ & 0.066 \\
\midrule
\emph{All engines and sectors} & +0.87 & +0.89 & +0.02 & $[-0.03, 0.07]$ & 0.484 \\
\bottomrule
\end{tabular}
\end{table}

\begin{table}[ht]
\centering\small
\caption{Study~1: slope of the normalised mention and citation rates on popularity, search on. Each rate is divided by its mean in the engine and sector, so a slope of 0.5 means that a brand searched ten times more often has a rate higher by half the average of its engine and sector. OLS with engine $\times$ sector fixed effects; 95\% intervals from 2,000 resamples of brands within each engine and sector.}
\label{tab:a_norm}
\begin{tabular}{l r r r r l}
\toprule
 & Brands & Mention & Citation & Difference & 95\% interval \\
\midrule
GPT & 702 & +0.52 & +0.51 & +0.02 & $[-0.08, 0.11]$ \\
Gemini & 702 & +0.56 & +0.18 & +0.38 & $[0.18, 0.58]$ \\
Perplexity & 702 & +0.57 & +0.83 & $-$0.26 & $[-0.44, -0.10]$ \\
\midrule
All engines & 2,106 & +0.55 & +0.50 & +0.04 & $[-0.05, 0.14]$ \\
\bottomrule
\end{tabular}
\end{table}

\begin{figure}[ht]
  \centering
  \includegraphics{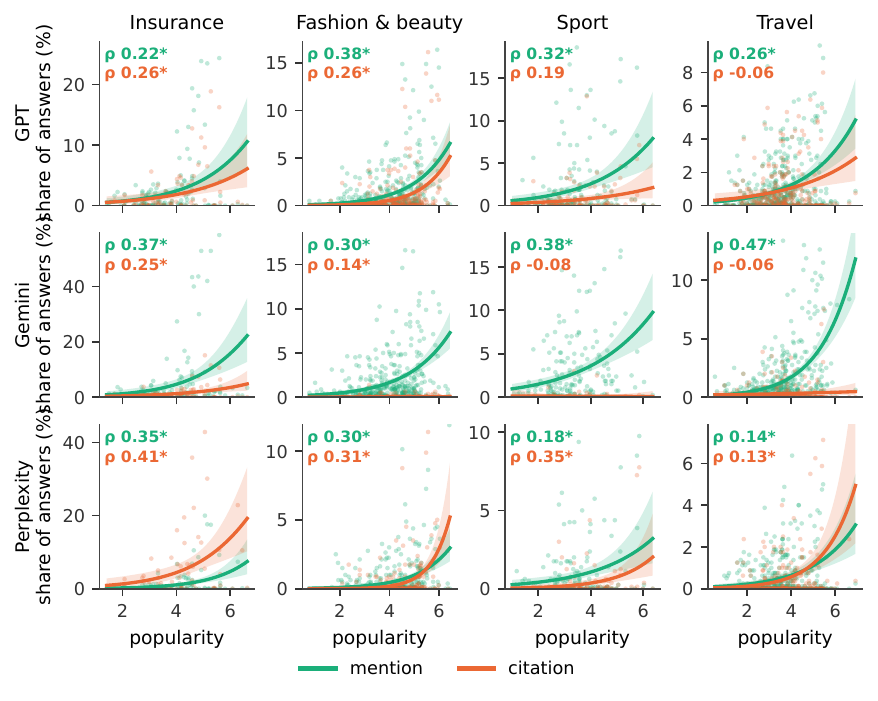}
  \Description{Grid of scatter plots, one per engine and sector, of mention and citation rates against popularity with fitted curves.}
  \caption{Mention and citation against popularity, one panel per engine and sector, search on. Points: brands (mention: all brands with a popularity value; citation: brands with a known domain). Lines: logistic fit of the share of answers on popularity, with 95\% band (quasi-binomial standard errors). In each panel, Spearman's $\rho$; $^{*}$ unadjusted $p < 0.05$ for Spearman's correlation, not for the fitted curve. Popularity is $\log_{10}(1+v)$, where $v$ is monthly Google search volume in Italy. The horizontal axis is shared across engines within each sector. The vertical limit is based on the 99.3rd percentile of the observed rates; the fits use all brands.}
  \label{fig:a_fits}
\end{figure}

\begin{figure}[ht]
  \centering
  \includegraphics{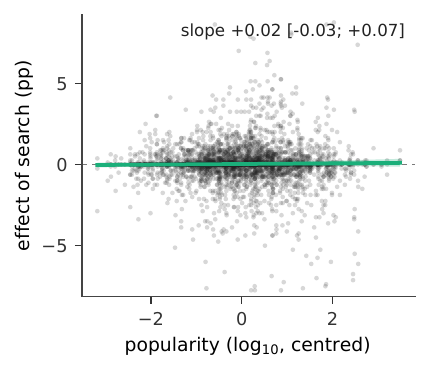}
  \Description{Scatter plot of the effect of search on each brand against popularity, with a flat fitted line near zero.}
  \caption{Effect of \texttt{search-on} on each brand (percentage points) against popularity ($\log_{10}$, centred within engine and sector), one point per brand and engine, with OLS line and 95\% bootstrap band over brands. 34 of 3,567 points fall outside the vertical range shown.}
  \label{fig:a_search_scatter}
\end{figure}

\subsection{Where the mentioned brands come from (H2)}
\label{app:h2}

Brand names in the search queries could make two answers that mention the same brands look similar in their search queries for that reason alone. We therefore remove from each search query the names of the brands mentioned for its sector and recompute query similarity. The slopes of Figure~\ref{fig:h2} remain positive: for sources 3.68 [3.27, 4.07] on GPT, 3.13 [2.76, 3.57] on Gemini and 4.24 [3.86, 4.68] on Perplexity (4.10, 3.42 and 4.59 with the original search queries), and for brands 3.05 [2.59, 3.52], 2.09 [1.69, 2.56] and 3.25 [2.64, 3.92] (3.57, 2.09 and 3.76).

\begin{figure}[ht]
  \centering
  \includegraphics{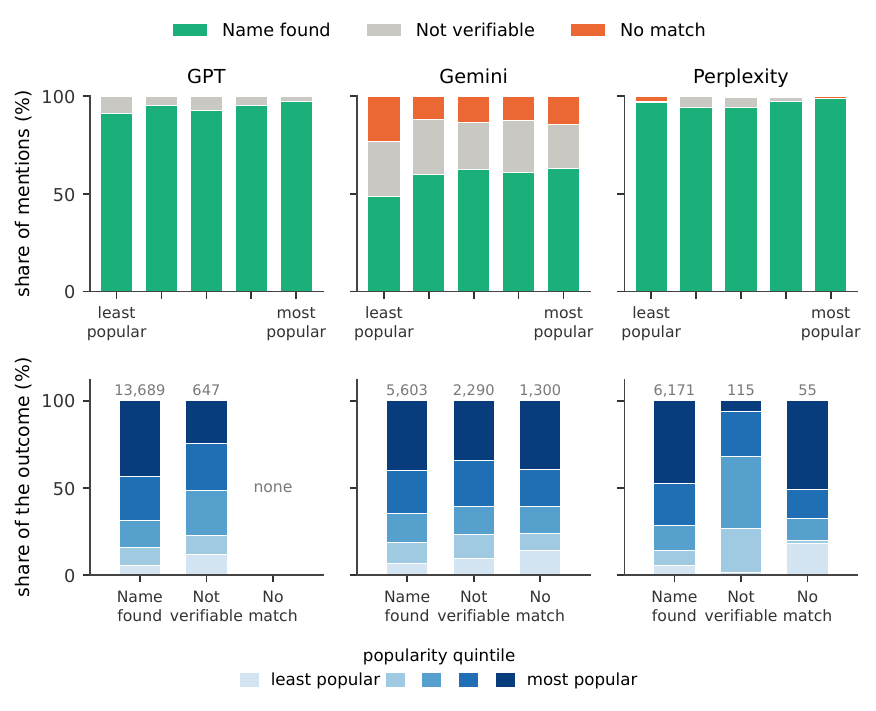}
  \Description{Stacked bar charts per engine. Top: share of mentions whose name is found in the returned sources, cannot be verified, or has no match, by quintile of popularity. Bottom: distribution of each outcome across quintiles.}
  \caption{Is the name of a mentioned brand in the sources the engine returned? Mentions in calls with observed search queries, for brands that reach the threshold of Section~\ref{sec:m_s1_metrics} in the engine itself, by quintile of popularity (brands split within engine and sector, as in Figure~\ref{fig:mention_citation}). Top: share of mentions whose name is found in the returned sources, that cannot be verified because some source could not be read, or with no match although every source was read. Bottom: the same mentions read the other way, how each outcome is distributed across the quintiles of popularity; numbers above the bars are mentions. GPT has no mention with no match. The match is lexical and the pages were read after the calls.}
  \label{fig:a_source}
\end{figure}

Between 34\% and 39\% of the brands GPT mentions appear only in its search actions after the first, which can react to what was found (Figure~\ref{fig:a_first_query}). The source check covers 30,006 mentions in calls with observed search queries. In 36.2\% of them the name of the brand is found only on third-party pages (GPT 29.1\%, Gemini 42.9\%, Perplexity 42.6\%). Mentions with no match do not become more frequent with popularity: the Spearman correlation is significant in 3 of the 8 engine--sector pairs of Gemini and Perplexity before correction, with opposite signs, and in none after Benjamini--Hochberg. On Gemini, 39.7\% of the mentions with no match concern brands in the most popular quintile and 14.0\% in the least popular, against 40.2\% and 6.8\% of the mentions whose name is found (Figure~\ref{fig:a_source}). A match does not show that the source caused the mention, and no match does not show that the mention came from memory, because we see only the sources the platform returns and we read the pages after the calls.


\subsection{Personas and the brand popularity bias (H3)}
\label{app:h3}

\begin{figure}[ht]
  \centering
  \includegraphics{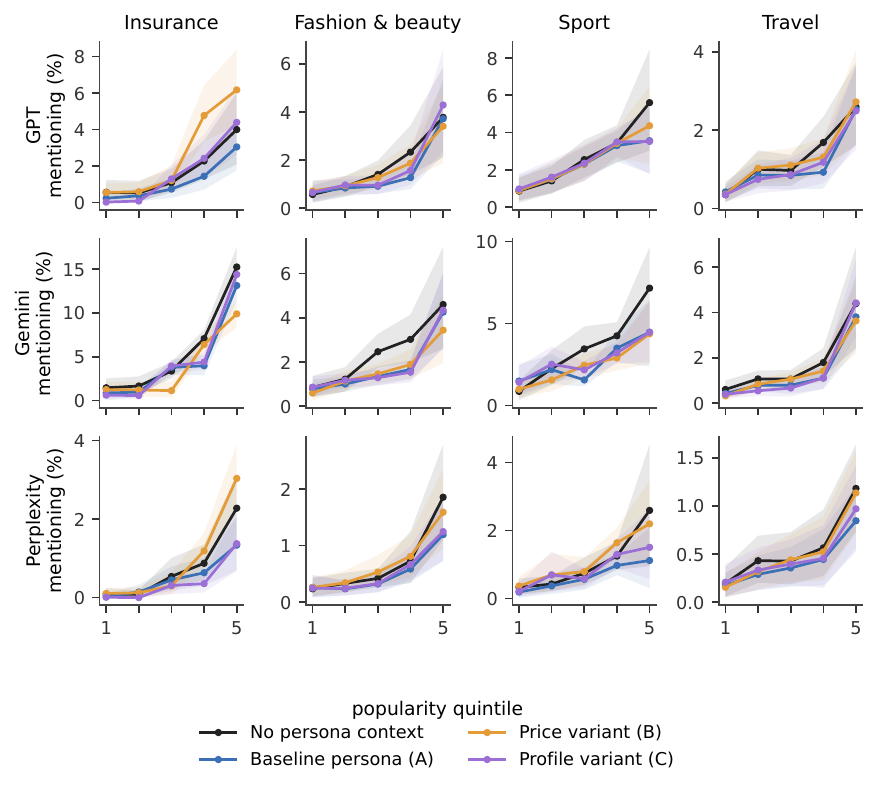}
  \Description{Grid of line charts, one per engine and sector, of the mention rate by quintile of popularity for no persona context and the three personas.}
  \caption{Mention rate by quintile of popularity and persona level, one panel per engine and sector, search off and on together; bands: 95\% intervals from resampling prompts.}
  \label{fig:a_persona}
\end{figure}

Without a persona context, a brand searched ten times more often on Google is mentioned in 0.98 more answers out of 100. The differences from this slope are $-$0.18 [$-$0.29, $-$0.07] with the Baseline persona (A), $-$0.13 [$-$0.24, $-$0.03] with the Price variant (B) and $-$0.10 [$-$0.21, 0.01] with the Profile variant (C). Figure~\ref{fig:a_persona} compares the three personas with no persona context, by engine and sector. In sport, with the Baseline persona (A), the slope of mention on popularity falls on GPT from 1.64 to 1.04, on Gemini from 2.02 to 1.19 and on Perplexity from 0.82 to 0.39. Other changes concern a single engine. On Gemini the sport slope falls with every persona. On Perplexity the insurance slope falls with the Baseline persona (A) and with the Profile variant (C), from 0.60 to 0.34 with both. Price variant (B) moves insurance in opposite directions, steeper on GPT (0.91 to 1.59) and flatter on Gemini (3.56 to 2.22). None of the 46 contrasts in the mean popularity of the mentioned brands between personas survives Benjamini--Hochberg, and those with $p < 0.05$ go in both directions. In insurance, the share of calls with a price word in the search queries rises from 4\% without a persona context to 37--40\% with the Baseline persona (A) or the Profile variant (C). With a persona, Gemini searches in 37--42\% of the calls, against 61\% without a persona context. The mean source Jaccard index between two repeats with the same persona is 0.39 on GPT, 0.23 on Gemini and 0.33 on Perplexity; between two answers with different personas it is lower by 0.074, 0.080 and 0.065, and by 0.030, 0.035 and 0.010 after adjusting for the similarity of the search queries.

\clearpage
\section{Study 2: questions, signatures and rewrites}
\label{app:s2}

\subsection{Choice of the names}
\label{app:s2_names}

The better-known brand is the candidate with the highest popularity among the brands most mentioned in the Study~1 answers to the prompt. The less-known brand is the least popular brand mentioned in those answers; it must not be one of the five pages, it must have a known official site, and it must not belong to the same group as another brand of the set. A page can be the target only if its popularity is at most half that of the better-known brand and at least twice that of the less-known one, so that the two other brands differ from the target by at least a factor of two in search volume. Invented names share no distinctive word, root, translation or form with the real ones, while generic terms, places, ingredients and facts stay as in the original.

\begingroup\scriptsize
\begin{longtable}{p{0.12\linewidth} p{0.33\linewidth} p{0.11\linewidth} p{0.11\linewidth} p{0.1\linewidth} p{0.08\linewidth}}
\caption{The 24 questions of Study~2 with the target page and its signatures. Better known and less known are other real brands mentioned in the Study~1 answers to the same prompt; the invented brand has no registered \texttt{.it} domain.}\label{tab:a_s2}\\
\toprule ID & Question & Target & Better known & Less known & Invented \\ \midrule \endfirsthead
\toprule ID & Question & Target & Better known & Less known & Invented \\ \midrule \endhead
\bottomrule \endlastfoot
assicurazioni\_03 & \textit{Quale assicurazione bici dovrei scegliere?} & AXA & Unipol & Qover & Dalveris \\
assicurazioni\_09 & \textit{Quale assicurazione sui voli mi consigli di scegliere?} & Heymondo & Genertel & ERGO Assicurazione Viaggi & Trevalda \\
assicurazioni\_10 & \textit{Qual è la migliore assicurazione natante da scegliere?} & Cattolica Assicurazioni & Allianz & Zurich & Corlenta \\
assicurazioni\_17 & \textit{Qual è l'assicurazione RCA migliore da scegliere?} & Allianz Direct & Prima Assicurazioni & Admiral & Corvalia \\
assicurazioni\_19 & \textit{Quale assicurazione spese veterinarie dovrei scegliere?} & ConTe.it & UniSalute & Baboop & Nidrava \\
assicurazioni\_20 & \textit{Quali sono le migliori opzioni per l'assicurazione vita in base ai costi?} & GenialLife & Unipol & SquareLife & Pelora \\
moda\_beauty\_03 & \textit{Quali profumi originali con prezzi bassissimi dovrei scegliere?} & Calvin Klein & Zara & Berdoues & Fiorvale \\
moda\_beauty\_05 & \textit{Qual è il miglior siero antimacchie da scegliere?} & The Ordinary & Caudalie & SkinCeuticals & Ilvessa \\
moda\_beauty\_08 & \textit{Quali sono i migliori jeans momfit da scegliere?} & OVS & Zara & Agolde & Velmora \\
moda\_beauty\_14 & \textit{Qual è la migliore crema contorno occhi da scegliere?} & Kiehl's & Shiseido & Neutrogena & Aurelys \\
moda\_beauty\_18 & \textit{Qual è il retinolo migliore?} & The Ordinary & La Roche-Posay & The Inkey List & Ondarys \\
moda\_beauty\_23 & \textit{Quali sono i profumi buoni da uomo che mi consiglieresti?} & Montblanc & Zara & Rochas & Nevrola \\
sport\_01 & \textit{Quali sono le migliori scarpe da trekking da scegliere?} & CMP & Adidas & Meindl & Dravio \\
sport\_09 & \textit{Qual è la migliore ebike per rapporto qualità prezzo?} & Trek & Specialized & Kalkhoff & Brivano \\
sport\_11 & \textit{Quali sono i migliori tapis roulant da scegliere?} & Everfit & Technogym & UREVO & Varenzo \\
sport\_13 & \textit{Qual è la racchetta da padel migliore da scegliere?} & Nox & Adidas & Wilson & Zorvane \\
sport\_17 & \textit{Quali sono le migliori cargo bike economiche da scegliere?} & Fiido & Engwe & Panda Bike & Brovale \\
sport\_21 & \textit{Quali sono le migliori scarpe da corsa da scegliere?} & ASICS & Nike & Mizuno & Kervino \\
viaggi\_04 & \textit{Quali villaggi scegliere per le vacanze in Cilento?} & Villaggio Elea & Baia del Silenzio & Camping Villaggio Saline & Ruvalia \\
viaggi\_07 & \textit{Quale villaggio turistico all inclusive mi consiglieresti?} & Club Esse & Alpitour World & Meraki Resort & Ombrevia \\
viaggi\_13 & \textit{Quali sono i migliori villaggi all inclusive alle Maldive del Salento?} & Nicolaus Club La Giurlita & Vivosa Apulia Resort & Arco del Saracino & Lunerio \\
viaggi\_15 & \textit{Quali sono i migliori tour operator per un viaggio a Zanzibar?} & Veratour & Alpitour World & Savannah Explorers & Calimbra \\
viaggi\_17 & \textit{Quali sono le migliori opzioni per il noleggio auto con prezzi bassi?} & Avis & Renault & Thrifty & Orbisa \\
viaggi\_18 & \textit{Quali sono i migliori tour operator siciliani da scegliere per un viaggio?} & Go-Etna & WeRoad & KidSicily & Brezzaria \\
\end{longtable}\endgroup

\begingroup\scriptsize
\begin{longtable}{p{0.13\linewidth} p{0.27\linewidth} p{0.27\linewidth} r r r}
\caption{Study~2, per question: the real products used for the better-known and less-known brands (-- where that signature has no real-product version), the distance in popularity between the target and each of the two real brands (orders of magnitude of monthly search volume: better known minus target, target minus less known), and how often the models name the true brand from the neutralised target page (12 answers per question).}\label{tab:a_s2_arms}\\
\toprule ID & Better known, real product & Less known, real product & Above & Below & Recognised \\ \midrule \endfirsthead
\toprule ID & Better known, real product & Less known, real product & Above & Below & Recognised \\ \midrule \endhead
\bottomrule \endlastfoot
assicurazioni\_03 & Unipol Unica Mobilità & Qover Qover-me & 0.41 & 2.58 & 8\% \\
assicurazioni\_09 & -- & -- & 0.34 & 1.88 & 50\% \\
assicurazioni\_10 & Allianz Columbus & Zurich Permare & 0.61 & 0.42 & 0\% \\
assicurazioni\_17 & -- & -- & 0.31 & 2.09 & 0\% \\
assicurazioni\_19 & -- & -- & 0.82 & 1.60 & 25\% \\
assicurazioni\_20 & Unipol Vita Smart & SquareLife Turtleneck & 1.80 & 0.77 & 0\% \\
moda\_beauty\_03 & Zara Red Temptation & Berdoues Monoï \& Tiaré & 1.23 & 3.35 & 50\% \\
moda\_beauty\_05 & Caudalie Vinoperfect Siero Illuminante Anti-Macchie & SkinCeuticals Discoloration Defense & 0.37 & 0.44 & 100\% \\
moda\_beauty\_08 & Zara Z1975 Mom Fit & Agolde 90's Pinch Waist & 0.42 & 2.89 & 33\% \\
moda\_beauty\_14 & Shiseido Benefiance Wrinkle Smoothing Eye Cream & Neutrogena Hydro Boost Contorno Occhi Anti-Fatica & 0.31 & 0.32 & 58\% \\
moda\_beauty\_18 & La Roche-Posay Retinol B3 Siero & The Inkey List Retinol Serum & 0.31 & 1.29 & 100\% \\
moda\_beauty\_23 & Zara Vibrant Leather & Rochas Moustache & 1.61 & 1.88 & 50\% \\
sport\_01 & Adidas Terrex AX4 & Meindl Bhutan MFS & 1.14 & 1.52 & 8\% \\
sport\_09 & Specialized Turbo Vado & Kalkhoff Endeavour & 1.07 & 1.07 & 92\% \\
sport\_11 & Technogym MyRun & UREVO Foldi 3S & 1.87 & 0.70 & 33\% \\
sport\_13 & Adidas Drive Black 3.5 & Wilson Blade V4 & 2.64 & 0.33 & 83\% \\
sport\_17 & Engwe LE20 & Panda Bike Cuore & 0.39 & 0.94 & 33\% \\
sport\_21 & Nike Pegasus 41 & Mizuno Wave Rider 29 & 0.71 & 0.60 & 67\% \\
viaggi\_04 & -- & -- & 0.89 & 1.77 & 25\% \\
viaggi\_07 & -- & -- & 1.69 & 1.36 & 17\% \\
viaggi\_13 & -- & -- & 0.74 & 0.39 & 0\% \\
viaggi\_15 & -- & -- & 0.58 & 2.42 & 17\% \\
viaggi\_17 & -- & -- & 0.37 & 1.25 & 8\% \\
viaggi\_18 & -- & -- & 2.55 & 1.92 & 50\% \\
\midrule
Mean & 15 questions & 15 questions & 0.97 & 1.41 & 38\% \\
\end{longtable}\endgroup

\subsection{Rewriting instructions (Italian, as sent to \texttt{gpt-5.5})}
Each rewrite instruction is followed by a common part.
\begin{itemize}\footnotesize
\item \textsc{Structure}: \textit{Riscrivi questo testo di una pagina prodotto per renderlo più chiaro e facile da usare per chi deve scegliere: riorganizzalo con brevi titoletti ed elenchi, metti prima le informazioni decisive, elimina ripetizioni.}
\item \textsc{Fluency}: \textit{Riscrivi questo testo per renderlo più scorrevole senza cambiarne il contenuto: le frasi devono legarsi bene l'una all'altra, la lingua deve essere chiara e coinvolgente e tutte le informazioni originali vanno conservate.}
\item \textsc{Authoritative}: \textit{Riscrivi questo testo in uno stile autorevole, senza aggiungere né togliere informazioni: deve trasmettere sicurezza e competenza, con affermazioni nette, e puoi rivolgerti al lettore in seconda persona. Mantieni la struttura (a capo, elenchi, ordine delle parti) e riformula solo singole frasi. Non aggiungere garanzie, primati o promesse.}
\item Common part: \textit{Il testo contiene segnaposto tra graffe come \{BRAND\} o \{TECH\_1\}: copiali identici, senza toglierne nessuno e senza aggiungerne. Usa SOLO le informazioni presenti nel testo: non aggiungere fatti, numeri, statistiche, citazioni, fonti, giudizi di terzi o promesse nuove, e non togliere nessuna informazione. Mantieni tutte le cifre esattamente come sono. Scrivi in italiano. Lunghezza simile all'originale: circa \{n\} parole, mai più di \{nmax\}. Restituisci solo il testo riscritto, in testo semplice (puoi usare a capo, trattini per gli elenchi e brevi titoletti seguiti da due punti; niente markdown come \# o **).}
\end{itemize}
\subsection{Ranking message (Italian, as sent to the four models)}
The message tells the model that the pages were found for the question and that their order is random. Each page begins with a line giving its domain and web address, followed by its text. {\footnotesize The question, then: \textit{Per rispondere usa le pagine qui sotto, trovate per questa domanda. L'ordine delle pagine è casuale.} Each page follows as \textit{[k] Fonte: dominio --- URL} and its text. The message ends with: \textit{Rispondi alla domanda usando queste pagine e citale con il loro numero tra parentesi quadre, per esempio [2]. Alla fine scrivi un'ultima riga con la tua classifica completa delle cinque pagine, dalla prima che consigli all'ultima, nel formato: CLASSIFICA: [n] > [n] > [n] > [n] > [n]}}

\subsection{Judging instruction (Italian, as sent to \texttt{deepseek-v3.2})}
{\footnotesize\textit{Confronta due versioni di un testo (i segnaposto tra graffe sono nomi). Conta come differenza SOLO un'informazione fattuale presente in una versione e del tutto assente nell'altra, oppure un valore cambiato (un numero, una caratteristica, una condizione, un prezzo, una copertura). NON sono differenze: titoli o intestazioni aggiunti o tolti, elenchi al posto di paragrafi, ordine diverso, sinonimi, riformulazioni con lo stesso significato, un riepilogo che ripete informazioni presenti altrove, punteggiatura, refusi corretti, frasi esortative o di puro tono. Elenca solo le differenze vere, una per riga, poi un'ultima riga 'ESITO: UGUALI' se non ce ne sono, altrimenti 'ESITO: DIVERSI'.}}

\subsection{Revised rewrites}
When a rewrite failed the checks three times, the attempt with the fewest errors was sent back to \texttt{gpt-5.5} with the original for a minimal correction of what the checks had found, and then went through the same checks and the same judge. This happened for the structure rewrite of 6 questions (assicurazioni\_03, assicurazioni\_09, assicurazioni\_19, sport\_17, viaggi\_04, viaggi\_17). The structure rewrite of sport\_21 was corrected by hand (final length 1.06 times the original) and passed the judge again. Every step is logged with the stimuli.

\section{Study 2: additional results}
\label{app:s2_results}

The effect of rewriting under each signature is +0.016 positions under the true brand, $-$0.001 with an invented product, +0.043 and +0.031 under the better-known and the less-known brand, and +0.023 under an invented brand. The differences between signatures are not significant (repeated-measures analysis of variance over the 24 questions), and the difference between the true and the invented brand is $-$0.007 [$-$0.057, 0.043].

\begin{table}[ht]
\centering\small
\caption{Study~2 by model and by sector: positions gained by the target page (positive = ranked higher), computed as in Table~\ref{tab:r2}. Each column is one row of that table, in the same order (\realdot{} real, \invdot{} invented product name, as in Figure~\ref{fig:r2}). Models average the 24 questions; sectors average six questions and the four models. Descriptive.}
\label{tab:a_s2_breakdown}
\begin{tabular}{l r r r r r}
\toprule
 & Original \realdot{} vs & Original \realdot{} vs & Better \invdot{} vs & Less \invdot{} vs & Rewrite vs \\
 & Invented \invdot & Original \invdot & Less \invdot & Invented \invdot & original \\
 & (H4a) & (H4b) & (H4c) & (H4e) & (H5a) \\
\midrule
DeepSeek & 0.24 & 0.12 & $-$0.06 & 0.16 & 0.05 \\
Gemini & 0.42 & 0.10 & 0.11 & 0.15 & 0.03 \\
GPT & 0.57 & 0.17 & 0.10 & 0.23 & 0.03 \\
Perplexity & 0.56 & 0.19 & 0.04 & 0.23 & $-$0.01 \\
\midrule
Insurance & 0.56 & 0.06 & 0.17 & 0.32 & 0.06 \\
Fashion and beauty & 0.68 & 0.39 & $-$0.02 & 0.11 & 0.03 \\
Sport & 0.26 & 0.11 & $-$0.05 & 0.17 & $-$0.01 \\
Travel & 0.27 & 0.02 & 0.09 & 0.18 & 0.01 \\
\midrule
All & 0.44 & 0.14 & 0.05 & 0.19 & 0.02 \\
\bottomrule
\end{tabular}
\end{table}

With the original text, the true brand gains more than 1.3 positions in three questions (Montblanc, Allianz Direct and Veratour), and leaving out any single question keeps the mean gain between 0.39 and 0.49. When the other brands carry a real product of their own instead of an invented one, the page rises by about 0.1 positions under either brand, which is not significant after correction. With their real products, the two brands rank the page above the invented brand (+0.27, Holm $p = 0.042$, and +0.29, Holm $p = 0.026$), do not differ from each other, and rank it lower than its true brand, though not significantly. Table~\ref{tab:a_s2_pairs} gives every pair of signatures.

\begin{table}[ht]
\centering\small
\caption{Study~2: every pair of the seven signatures (\realdot{} real, \invdot{} invented product name, as in Figure~\ref{fig:r2}). $\Delta$: positions gained by the target under the first signature over the second, averaged over the four texts; 95\% intervals from resampling questions. Pairs with a real-product version of \emph{Better known} or \emph{Less known} use the 15 questions where it exists. Family: H4a--H4e are rows of Table~\ref{tab:r2} (one-sided $p$, Holm over its seven rows); R is the family of seven comparisons with the real-product versions (two-sided $p$, Holm over the seven); the other pairs are exploratory (two-sided $p$, not adjusted).}
\label{tab:a_s2_pairs}
\begin{tabular}{l l r l r c c}
\toprule
\textbf{First} & \textbf{Second} & $\boldsymbol{\Delta}$ & \textbf{95\% interval} & \textbf{Questions} & \textbf{Family} & $p$ \\
\midrule
Original \realdot & Original \invdot & $+0.14$ & $[0.05, 0.27]$ & 24 & H4b & $<.05$ \\
Original \realdot & Better known \realdot & $+0.17$ & $[-0.06, 0.45]$ & 15 & R & n.s. \\
Original \realdot & Better known \invdot & $+0.20$ & $[0.04, 0.39]$ & 24 & -- & $<.05$ \\
Original \realdot & Less known \realdot & $+0.15$ & $[-0.01, 0.33]$ & 15 & R & n.s. \\
Original \realdot & Less known \invdot & $+0.25$ & $[0.05, 0.47]$ & 24 & -- & $<.05$ \\
Original \realdot & Invented \invdot & $+0.44$ & $[0.23, 0.69]$ & 24 & H4a & $<.01$ \\
Original \invdot & Better known \realdot & $-0.04$ & $[-0.17, 0.08]$ & 15 & -- & n.s. \\
Original \invdot & Better known \invdot & $+0.06$ & $[-0.04, 0.15]$ & 24 & -- & n.s. \\
Original \invdot & Less known \realdot & $-0.06$ & $[-0.15, 0.02]$ & 15 & -- & n.s. \\
Original \invdot & Less known \invdot & $+0.11$ & $[-0.06, 0.29]$ & 24 & -- & n.s. \\
Original \invdot & Invented \invdot & $+0.30$ & $[0.13, 0.51]$ & 24 & -- & $<.01$ \\
Better known \realdot & Better known \invdot & $+0.11$ & $[0.03, 0.21]$ & 15 & R & n.s. \\
Better known \realdot & Less known \realdot & $-0.02$ & $[-0.16, 0.11]$ & 15 & R & n.s. \\
Better known \realdot & Less known \invdot & $+0.08$ & $[-0.03, 0.20]$ & 15 & -- & n.s. \\
Better known \realdot & Invented \invdot & $+0.27$ & $[0.12, 0.44]$ & 15 & R & $<.05$ \\
Better known \invdot & Less known \realdot & $-0.13$ & $[-0.27, -0.02]$ & 15 & -- & n.s. \\
Better known \invdot & Less known \invdot & $+0.05$ & $[-0.11, 0.22]$ & 24 & H4c & n.s. \\
Better known \invdot & Invented \invdot & $+0.24$ & $[0.08, 0.44]$ & 24 & -- & $<.05$ \\
Less known \realdot & Less known \invdot & $+0.10$ & $[0.02, 0.19]$ & 15 & R & n.s. \\
Less known \realdot & Invented \invdot & $+0.29$ & $[0.14, 0.46]$ & 15 & R & $<.05$ \\
Less known \invdot & Invented \invdot & $+0.19$ & $[0.11, 0.29]$ & 24 & H4e & $<.01$ \\
\bottomrule
\end{tabular}
\end{table}

Asked which brand a page with all names replaced belongs to, the models name the true brand in 38\% of the answers, an upper bound since the question invites a guess, from none in four questions to every answer for the two pages of The Ordinary (Table~\ref{tab:a_s2_arms}); Gemini recognises the brand most often (65\% of its answers) and DeepSeek least often (14\%). In the rankings, no answer mentions the original brand when the page carries another signature, and the effect of the true brand is not related to how often the page is recognised (Spearman $\rho = 0.10$). The effect of the product name (H4b), instead, comes mostly from the pages the models recognise: it is 0.03 positions on the 11 questions recognised at most a quarter of the time and 0.24 on the other 13. Of the three subsets of questions fixed before the collection, the true brand gains 0.48 positions on the 20 questions whose target ranked between 2 and 4 in the calibration run, 0.36 on the 14 questions where the models rate the better-known brand as better known than the target, and 0.40 when we drop the four questions where the brand appears only in the web address. On the same 14 questions, H4c is +0.15, not significant. The balanced orders of the pages cancel a position effect: the target ranks 3.27 when shown first and 3.54 when shown last.


\section{Summary of hypotheses}
\label{app:hyp}

\begin{table}[ht]
\centering
\small
\caption{Hypotheses, predictions and outcomes, by research question. Gains in Study~2 are positions on a scale of 1 to 5.}
\label{tab:hyp_summary}
\begin{tabularx}{\linewidth}{l X X}
\toprule
 & \textbf{Prediction} & \textbf{Outcome} \\
\midrule
\multicolumn{3}{l}{\emph{RQ1, Study 1: brand popularity bias}} \\
\HH{1a} & More popular brands are mentioned more often & Supported in 23 of 24 configurations \\
\HH{1b} & More popular brands are cited more often & Partially supported: 8 of 12 engine--sector pairs \\
\HH{1c} & The brand popularity bias is stronger with \texttt{search-off} than \texttt{search-on} & Not supported (+0.02 [$-$0.03, 0.07]) \\
\HH{2a} & Answers with more similar search queries share more sources and text & Supported \\
\HH{2b} & Their mentioned brands do not overlap more & Not supported: brands follow the search queries almost as closely as sources \\
\HH{3a} & The brand popularity bias is the same with and without a persona & Not supported for the Baseline persona (A) and the Price variant (B) \\
\HH{3b} & Answers with different personas are as similar as answers with the same persona & Not supported: a gap remains at equal query similarity \\
\midrule
\multicolumn{3}{l}{\emph{RQ2, Study 2: brand familiarity bias (H4)}} \\
\HH{4a} & A page ranks higher under its real brand than under an invented one & Supported (+0.44) \\
\HH{4b} & It ranks higher with its real product name than with an invented one & Supported (+0.14) \\
\HH{4c} & A more popular real brand ranks it higher than a less popular one & Not supported (+0.05) \\
\HH{4d} & Another real brand lowers it because it does not match the page & Not significant (+0.08) \\
\HH{4e} & A less popular real brand ranks it higher than an invented one & Supported (+0.19) \\
\midrule
\multicolumn{3}{l}{\emph{RQ2, Study 2: rewriting (H5)}} \\
\HH{5a} & A rewrite raises its rank & Not supported (+0.02) \\
\HH{5b} & The real brand gains more than rewriting & Supported (+0.42) \\
\bottomrule
\end{tabularx}
\end{table}

\end{document}